\documentstyle[12pt,epsf,aaspp4]{article}
\font\smcap=cmcsc10
\def\Catwo{Ca$\>${\smcap ii}}
\def\farcs{\hbox{$\> .\!\!^{\prime\prime}$}}

\def\fss{\hbox{$\> .\!\!^{\rm s}$}}
\def\refpar{\par\hangindent=3em\hangafter=1}
\def\reference{\relax\refpar}

\begin{document}
\singlespace

\title{Isolating Red Giant Stars in M31's Elusive Outer Spheroid}

\author{David B.\ Reitzel and Puragra Guhathakurta\altaffilmark{1}}
\affil{UCO/Lick Observatory, University of California, Santa Cruz, California
   95064, USA}
\affil{Electronic mail: {\tt reitzel@ucolick.org}, {\tt raja@ucolick.org}}
\authoremail{reitzel@ucolick.org, raja@ucolick.org}

\author{Andrew Gould\altaffilmark{1}}
\affil{Department of Astronomy, Ohio State University, 174 West 18th Avenue,
   \\
   Columbus, Ohio 43210-1106, USA}
\affil{Electronic mail: \tt gould@astronomy.ohio-state.edu}
\authoremail{gould@payne.mps.ohio-state.edu}

\altaffiltext{1}{Alfred P.\ Sloan Research Fellow}

\begin{abstract}

\noindent
Deep $UBRI$ images of a $15'\times15'$ field in the outer spheroid of the M31
galaxy obtained using the Kitt Peak National Observatory 4-meter telescope,
and $I$ band images obtained with the Keck 10-meter telescope and Low
Resolution Imaging Spectrograph, are used to isolate a sample of candidate
red giant branch stars located at a projected radial distance of 19~kpc along
the minor axis.  These stars are distinguished from the more numerous distant
field galaxies on the basis of broadband $U-B$, $B-R$, and $R-I$ colors and
image morphology: We isolate objects whose colors are consistent with the
long, but relatively narrow locus occupied by red giants in $UBRI$ color
space (as observed in Galactic globular cluster giants and predicted by
models spanning a wide range of metallicities and ages), and those whose
angular sizes are consistent with the stellar point spread function, $\rm
FWHM=0\farcs6$--$0\farcs9$ (Keck) and $\rm FWHM=0\farcs9$--$1\farcs5$ (Kitt
Peak).  We carry out the same analysis of data on a comparison field with a
similar Galactic latitude to the M31 halo field.  The color-magnitude diagram
of objects in the comparison field is well described by a superposition of
foreground Galactic dwarf stars (in keeping with a standard empirical model
of the Galaxy) against a backdrop of contaminating faint blue field galaxies
($I\ga21$, $B-I\sim1\>$--$\>$2.5), while the M31 halo field contains a clear
excess of faint red objects ($I\sim20\>$--$\>$23, $B-I\sim2\>$--$\>$3.5) in
addition to these two components.  The location of this population of faint
red objects in the color-magnitude diagram is as would be expected for red
giant stars at the distance of M31.  The surface density of red giant
candidates in the $R=19$~kpc M31 halo field is consistent with the
findings of two recent {\it Hubble Space Telescope\/} studies.  The
data indicate that M31's stellar halo is much denser and/or larger
than that of the Galaxy: $(\rho_{\rm M31}^{\rm  RGB}/\rho_{\rm
MW}^{\rm RGB})(\Lambda/1.5)^{-\nu}\sim10$, where $\Lambda$ is the
ratio of the radial scale lengths of M31 and the Galaxy and $\nu=-3.8$
is the assumed power law index of the density profile; in fact, M31's
profile may be steeper than this ($\nu<-3.8$).  The color and slope of
the red giant branch in M31's outer halo are suggestive of a
relatively metal-rich population, $\rm[Fe/H]\ga-1$, in agreement with
the {\it Hubble Space Telescope\/} measurements.

\end{abstract}

\keywords{galaxies: individual: Andromeda galaxy [Messier~31 (M31), NGC~224,
UGC~454, CGCG~535-017] -- galaxies: formation -- stars: red giants -- stars:
metallicity}

\section{Introduction}

Studying the metallicity gradient of galactic spheroids is crucial for
understanding their formation and evolutionary history.  The dissipational
collapse model (Eggen et~al.\ 1962; Larson 1974) predicts a strong
metallicity gradient because of progressive chemical enrichment during the
collapse of the protogalactic gas cloud.  The accretion model (Searle \& Zinn
1978), on the other hand, predicts no strong gradient because star formation
largely precedes assembly of the galactic spheroid.  The metallicity gradient
of a halo can be measured in both the globular cluster system and in field
stars; in fact, it is important to study both populations independently since
they may be dynamically distinct from each other.

Globular clusters are easily observed out to large distances in the Galaxy
because of their high luminosity and this allows a reliable determination of
the metallicity gradient in the system of Galactic globular clusters.  Zinn
(1993) finds evidence for two subsystems among clusters: an `old halo'
population which displays a radial [Fe/H] gradient, a small age spread, and
significant rotation, and a `young halo' population which displays no
metallicity gradient, a large age spread, and very little rotation.  He
associates the former population with the initial dissipational collapse of
the Galactic halo (a la Eggen et~al.\ 1962) and the latter with subsequent
accretion of stellar systems (a la Searle \& Zinn 1978).  On the other hand,
Carney et~al.\ (1990) find a mean metallicity of $\rm[Fe/H]=-1.72$ and an
absence of a strong metallicity gradient as a function of apogalacticon
distance over the range $8~{\rm kpc}\leq{R}_{\rm apo}\leq40$~kpc in a proper
motion selected survey of 227~Galactic field halo stars with velocities
lagging the local standard of rest's circular velocity by at least
150~km~s$^{-1}$.

It seems logical to test the universality of the
conclusions about galaxy formation which are being drawn from the single case
of the Galaxy.  The Andromeda galaxy (M31) provides an external perspective
of a large spiral similar to our own and yet is close enough for individual
stars to be resolved.  Huchra et~al.\ (1991) showed that there is evidence
for a weak metallicity gradient in a sample of 150~M31 globular clusters,
with a mean metallicity of $\rm[Fe/H]=-1.2$, which is slightly higher than
the mean value of $\rm[Fe/H]=-1.4$ for Galactic globular clusters.

Over the last decade, several groups have tried to determine the metallicity
of field red giant branch (RGB) stars in M31's spheroid, starting with the
early work of Crotts (1986).  Mould \& Kristian (1986) found a metallicity of
$\rm[Fe/H]=-0.6$ for stars in a field located at a projected distance of
$R=7$~kpc along the minor axis, and assumed that the observed color
spread was due to a large 
metallicity dispersion.  Mould (1986) failed to determine the metallicity at
$R=20$~kpc, because the contaminating field galaxies could not be removed by
the standard method of morphological selection and statistical subtraction.
Pritchet \& van den Bergh (1988) estimated $\rm[Fe/H]=-1.0$ at $R=8.6$~kpc on
the minor axis, and again attributed the large color spread to an intrinsic
spread in [Fe/H].  Christian \& Heasley's (1991) data in an $R=16$~kpc field
around the globular cluster G219 suggested a large metallicity spread, with a
mean value $\rm[Fe/H]\ga-1.0$, as did Davidge's (1993) study at $R=6.7$~kpc,
Durrell et~al.'s (1994) study, and Couture et~al.'s (1995) study which
targeted fields around five~M31 globular clusters.  While these studies have
furthered our knowledge of M31's halo, sample contamination by distant field
galaxies, foreground Galactic dwarfs, and M31 disk giants poses a serious
obstacle in attempts to determine the metallicity of M31's outer halo
($R\ga10$~kpc).

\begin{figure}
\epsscale{0.8}
\plotfiddle{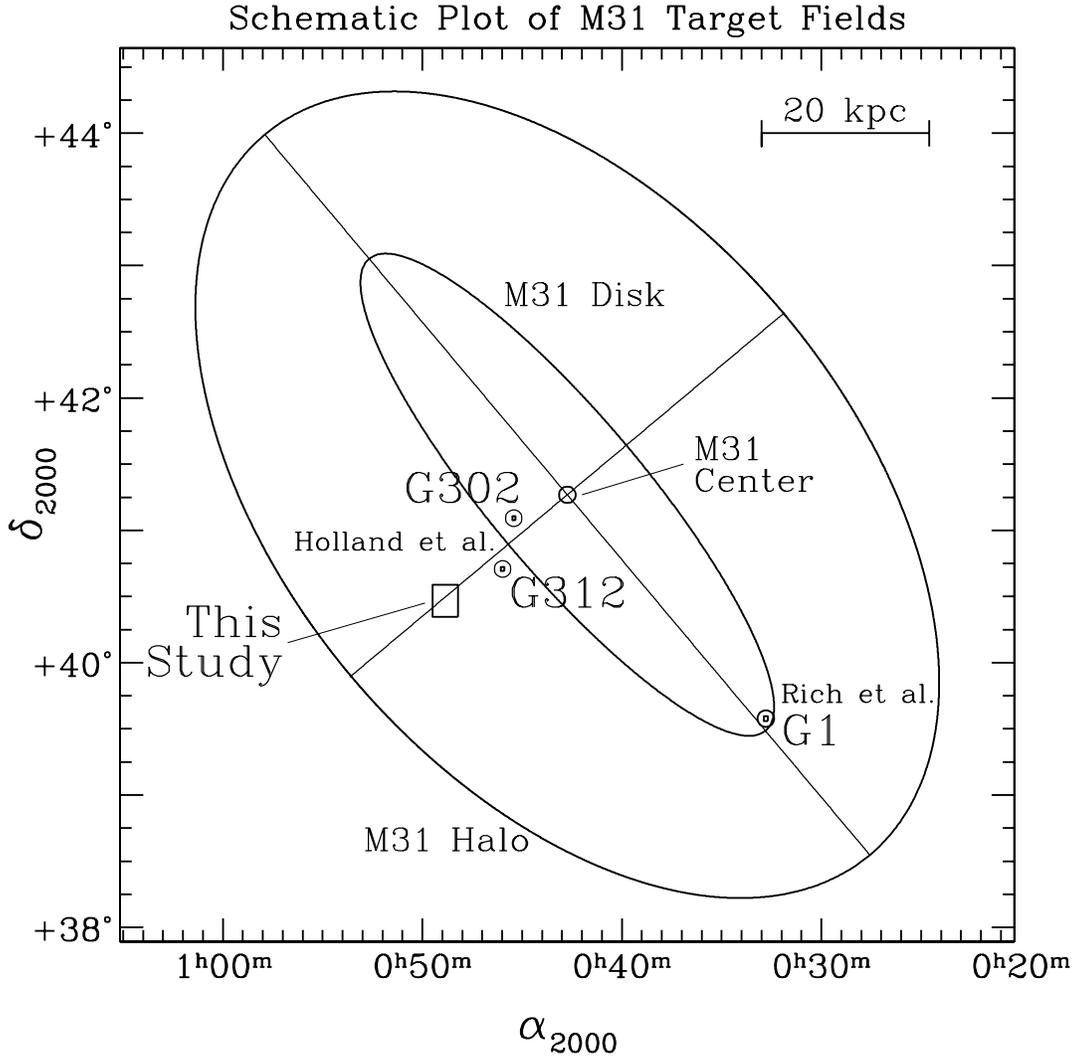}{4in}{0}{70}{70}{-210}{-90} 
\caption{Schematic plot showing the approximate
location of the field studied in this paper, along with Holland
et~al.'s (1996) G302 and G312 fields and Rich et~al.'s (1996) G1
field.  The size of each square symbol marking field position
corresponds to the actual area of the field studied, but not to the
field shape.  The ellipses indicate the orientation of M31's disk
($i=77^\circ$, $\rm PA=40^\circ$, with finite thickness) and halo (5:3
flattening, same PA as disk); the sizes of the ellipses are arbitrary.
The 10\% change in the $\rm cos(\delta)$ projection factor from the
top to the bottom of the plot has been ignored in this schematic
representation. \label{fig1}}
\end{figure}

Two recent studies of the M31 spheroid have taken advantage of the excellent
angular resolution of {\it Hubble Space Telescope\/} ({\it HST\/}) Wide
Field/Planetary Camera~2 images to separate stars from distant field
galaxies.  In the first study, Rich et~al.\ (1996) analyze archival {\it
HST\/} images of a field centered on the M31 globular cluster G1.  The line
of sight to this field (Fig.~\ref{fig1}) intersects M31's highly inclined
disk ($i\sim77^\circ$; de~Vaucouleurs 1958) at a relatively large radius
$R_{\rm disk}\approx{R}=40$~kpc on the major axis
(the value of $R_{\rm disk}$ could be different if M31's outer disk is
warped).  The second study is that of Holland et~al.\ (1996) who analyze deep
{\it HST\/} images of two~fields around M31 globular clusters located
7.6~kpc (G302) and 10.8~kpc (G312) from the center roughly along the SE minor
axis (Fig.~\ref{fig1}).  Both studies find the mean 
metallicity of M31's halo RGB stars to be comparable to that of 47~Tuc
($\rm[Fe/H]=-0.7$), with a spread of nearly 2~dex, suggesting that there was
a greater degree of pre-enrichment during the assembly of M31's spheroid than
in the case of the Galaxy's spheroid.  We will return to a more detailed
discussion of these studies later in the paper (Sec.~4.2).

In this paper, we probe RGB stars in the M31 halo using an alternative
method for discriminating between stars and faint galaxies; one that uses a
combination of broadband $UBRI$ colors and morphological information (Gould
et~al.\ 1992, hereafter referred to as GGRF).  Early results have been
reported by Reitzel et~al.\ (1996a,b).  The field of study presented here is
at a projected distance of $R=19$~kpc from M31's center along the minor axis
(only Rich et~al.'s G1 field is at a comparable radial distance out in the
spheroid assuming it has a 5:3 flattening) and the line of sight to the field
intersects M31's disk at $R_{\rm disk}\approx75$~kpc, further out than in any
of the previous studies (Fig.~\ref{fig1}).  The area of the field in this
study, 210~arcmin$^2$, is more than 60~times larger than each of the
Holland et~al.\ fields, and more than 40~times larger than the Rich et~al.\
G1 field.  Moreover, the field stars targeted in this study are well away
from any of M31's globular clusters.  The observations and data reduction
procedure are described in Sec.~2.  The technique  used to isolate M31 halo
RGB stars from background field galaxies and foreground Galactic dwarfs is
described in Sec.~3.  In Sec.~4, we discuss the properties of M31's spheroid
in the context of the recent {\it HST\/} studies and outline plans for future
work.  The main points of the paper are summarized in Sec.~5.

\section{The Data}
\subsection{Observations}

We present observations of an outer spheroid field in M31 and a comparison
field carried out with the the Kitt Peak National Observatory (KPNO) 4-meter
telescope and a $2048\times2048$ Tektronix CCD during 4~nights in
November~1993.  The M31 halo field:

\begin{equation}
\rm\alpha_{J2000}\>(M31~halo) =  00^h 48^m  51\fss 0;~~
   \delta_{J2000}\>(M31~halo) = +40^\circ 28' 05''
\end{equation}

\noindent
is located $84'$ from the galaxy's center, at a projected radial distance of
18.8~kpc for an assumed distance modulus of $(m-M)_0=24.43$~mag or
$D=770$~kpc (Freedman \& Madore 1990), at a position angle of $124.8^\circ$
(E of N) putting it close to M31's SE minor axis (Fig.~\ref{fig1}); the
position angle of the M31 disk major axis is $40^\circ$.  The comparison
field:

\begin{equation}
\rm\alpha_{J2000}\>(Comp) =  07^h 49^m 00\fss 0;~~
   \delta_{J2000}\>(Comp) = +60^\circ 03' 46''
\end{equation}

\noindent
provides a control sample: its Galactic
coordinates [$l^{II}({\rm Comp})=157^\circ$, $b^{II}({\rm Comp})=+30^\circ$]
are roughly similar to those of the M31 halo field [$l^{II}({\rm
M31~halo})=122^\circ$, $b^{II}({\rm M31~halo})=-22^\circ$].  We compare
the number, magnitude distribution, and color distribution of Galactic
stars in the two fields in the context of the Bahcall \& Soneira (1984) model
(see Sec.~3.4 for details).

The data set consists of images through $U$, $B_J$, $R$, and $I$ filters with
bandpasses centered at 3600, 4500, 6500, and 9000\,\AA, respectively.  Note,
this bandpass system is roughly similar, but {\it not\/} identical, to the
standard Johnson/Kron/Cousins bandpass system.  The pixel scale is
$0\farcs47$ and the field of view of each individual CCD frame is
$16'\times16'$.  The FWHM of the seeing disk ranges from
$0\farcs9\>$--$\>1\farcs5$ across the data set, and is generally smallest for
$I$ band exposures.  The primary observations consist of short (15~min in
$U$, 10~min in each of $B_JRI$), slightly disregistered exposures of the M31
halo field: a total of 21~exposures in $U$, 14~in $B_J$, 13~in $R$, and 14~in
$I$ corresponding to effective total exposure times of 5.3~hr in $U$ and
2.2$\>$--$\>$2.3~hr in each of $B_JRI$.  The total exposure times are
somewhat shorter for the comparison field: 15~exposures in $U$ (3.8~hr),
11~in $B_J$ (1.8~hr), 8~in $R$ (1.3~hr), and 9~in $I$ (1.5~hr).

In order to empirically verify the locus of RGB stars in multicolor space,
we use short $UB_JRI$ exposures of 6~Galactic globular clusters spanning a
range of metallicities.  The data for the clusters M15, M79, and M2 are of
adequate depth and quality; we are unable to use the data for the other
3~clusters, NGC~7006, NGC~6934, and Pal~11, as the surface density of cluster
giants is too low relative to contaminating field Galactic dwarfs.  Short
$UB_JRI$ exposures of standard star fields (Landolt 1983) are used for
photometric calibration of the data.  The globular cluster and standard star
exposures are from two~photometric nights during the observing run.  In
addition, a sequence of zero-second bias frames and short $I$ band twilight
flat exposures are used for data processing.

For the purpose of measuring angular sizes of faint objects, we use $I$ band
images of the M31 halo field obtained using the Keck 10-meter telescope and
the Low Resolution Imaging Spectrograph (LRIS, Oke et~al.\ 1995) during
1~night in September~1995.  The 4~Keck/LRIS exposures, $1\times10$~min and
$3\times5$~min, form a (mostly) non-overlapping $2\times2$ mosaic covering
about 75\% of the field of view of the KPNO image.  The scale of the Keck
images is $0\farcs22$~pixel$^{-1}$ and the field of view of each CCD frame is
$5\farcm7\times7\farcm3$.  The FWHM of seeing is $0\farcs6\>$--$\>0\farcs9$
for the Keck observations.  Limited use is made of these Keck data; unless
specifically mentioned, all subsequent data references in the paper pertain
to the main KPNO data set. 

\subsection{Flat Fielding, Fringe Removal, \& Coaddition of the Data}

Each CCD image is overscan subtracted, trimmed, and bias subtracted in the
usual way.  The $U$, $B_J$, and $R$ images are then flat fielded using ``dark
sky flats''.  Each dark sky flat is the median combination, with $\pm2\sigma$
rejection of cosmic rays, stars, and galaxies, of all the disregistered
images in that band of both M31 halo and comparison fields
(20$\>$--$\>$35~images per filter).  The $I$ band images are flat fielded
differently from those in $UB_JR$ as the raw CCD frames in $I$ contain
Fabry-Perot fringing of night sky emission lines.  An $I$ band flat field
image is created by median combining twilight flat exposures with
$\pm2\sigma$ rejection; these images are dominated by the (solar) continuum
and are largely free of fringing.  All $I$ exposures are first flat fielded
with the twilight flat, which corrects for spatial variation of quantum
efficiency across the CCD but leaves the fringe pattern intact.  A fringe
template is created by median combining the twilight-flat-fielded $I$
exposures of the M31 halo and comparison fields with $\pm2\sigma$ rejection,
and by reducing the resulting image to zero mean.  The amplitude of the
fringe pattern is found to vary from image to image (roughly
0.5\%$\>$--$\>$1.5\% of the sky level), so the fringe template is scaled
individually to optimize fringe removal for each $I$ band image. 

The images are geometrically transformed to correct for slight image
distortion ($\la10^{-3}$) and aligned for coaddition.  Each image is scaled
to correct for any variation in transparency as a function of airmass and/or
time (typically $\la10\%$), as determined from measurements of secondary
photometric standard stars on the image.  An additive offset is then applied
to match the sky background level over common areas of all images in a given
band, correcting for time variations in the night sky brightness (most
extreme in the $I$ band).  The transparency- and background-matched images in
each band are median combined with $\pm2\sigma$ rejection of cosmic rays.
Cosmic rays are then masked from each image by comparing the image to the
median image in that bandpass.  Finally, all images in a given band
($UB_JRI$) are averaged, excluding masked pixels.  The combined images each
cover an area of about $15'\times15'$; the dithering between exposures
slightly reduces the overlap area compared to the field of view of individual
CCD frames.

\subsection{Artifact Correction}

The combined CCD images suffer from two kinds of artifacts in the vicinity of
bright stars ($B\la17$, $I\la15$): charge bleeding and amplifier hysteresis.
These problems are most severe in the $I$ band image (as the brightest stars
are quite red in color and because the instrumental throughput is highest in
this band) and least in the $U$ band image.  While the artifacts cover only a
small fraction of the total image area ($\sim1\%$), they hamper the
performance of the automated object detection and photometry software over a
larger surrounding area.  We apply a set of techniques that reduce, but do
not eliminate, the effect of artifacts in the M31 halo field image.  We do
not correct artifacts in the comparison field as it is less severely affected
than the M31 halo field (higher Galactic latitude, and hence fewer bright
stars).  The affected areas of the comparison field ($\approx6\%$ of its
area), and the artifact-corrected M31 halo field ($\approx8\%$ of its area)
are properly accounted for in the subsequent analysis (Sec.~2.5).

Extreme saturation of pixels near the center of the image of a bright star
can lead to the overflow of charge along CCD columns ($y$ axis).  We
interpolate the sky background row by row across the affected region
by fitting a linear function in $x$ with a $\pm2\sigma$ rejection to
avoid stars/galaxies.  While this procedure obviously does not  
recover any of the data in the pixels occupied by the bleed, it replaces each
bleed with a smoothly varying background, thereby allowing faint objects to
be detected in the neighborhood of bright stars.

Amplifier hysteresis results in a temporary change in the effective gain
immediately after reading out saturated pixels ($\sim1$\% decrease in $R$ and
$I$, $\la10$\% increase in $B$ and $U$), with the gain gradually returning to
its nominal level.  This produces prominent streaks of low or high values
extending for several hundred pixels ($2'\>$--$\>6'$) along the $x$ axis to
the right of all bright stars (i.e.,~downstream in the readout sequence).
Multiple streaks caused by the numerous bright stars in the M31 halo field
combine to form an uneven background and affect the probability of faint
object detection.  Each of these streaks is rectified as follows:
(1)~A linear function in $x$, $f_{\rm ref}(x)$, is fit to a 10~pixel-wide (in
  $y$) reference region above and below the streak, extending along the full
  length of the streak in $x$, with $\pm2\sigma$ rejection of stars/galaxies; 
(2)~A linear function in $x$, $f_i(x)$, is similarly fit to each row $i$ of
  the streak; and
(3)~The difference, $f_{\rm ref}(x)-f_i(x)$, is added to each row ($i$) of
  pixels in the streak.
This technique does an excellent job of correcting for amplifier hysteresis
while preserving: (a)~the intensity distribution of a star or galaxy whose
image happens to lie, partly or wholly, within a streak and (b)~possible
large scale gradients in the sky background level.

\subsection{Detection and Photometry of Stars and Galaxies}

After rectifying charge bleed and amplifier hysteresis artifacts in the data,
we construct a single deep image for each of the M31 halo and comparison
fields.  This image is constructed from the weighted sum of the images
in all 4~bands.  The weights are
chosen to maximize the signal-to-noise ratio for a faint object whose
$UB_JRI$ colors are roughly equal to those of a typical faint RGB star.  The
software package {\smcap focas} (Jarvis \& Tyson 1981; Valdes 1982) is used
to analyze the M31 halo field and comparison field data in the following way:
(1)~A catalog of individual objects (stars and galaxies) is derived from the
  deep, summed $U$+$B_J$+$R$+$I$ image using a matched filter search
  technique (with search parameters tuned to maximize faint object detection
  efficiency while minimizing the fraction of spurious detections), and a
  limiting isophote is defined for each object; and
(2)~For each star/galaxy, this limiting isophote is applied to the image in
  each of the 4~bands ($UB_JRI$) to obtain photometry---i.e.,~color
  measurements are based on these ``common optimized apertures''.
We have checked that, for a set of relatively isolated objects, the {\smcap
focas} apparent magnitude agrees to within $\la0.03$~mag with the ``total'' 
apparent magnitude derived from a standard curve of growth analysis.

After accounting for the difference in line-of-sight reddening between the
M31 halo and comparison fields, $E(B-V)=0.085$ and 0.038, respectively
(Burstein \& Heiles 1982), we find a small offset between the $UBRI$ loci of
bright (foreground) stars between the two fields corresponding to a residual
magnitude zeropoint difference in the $B$ band.  This difference is probably
caused by errors in the transparency correction; empirical checks indicate
that the difference is not likely to be a result of PSF differences or errors
in the assumed $E(B-V)$.  We apply a slight adjustment to the $B$ magnitude
scales of both data sets ($0.1$~mag) to bring them into agreement with each
other and with the theoretical Bertelli et~al.\ (1994) $UBRI$ stellar loci
(Sec.~3.2).  This zeropoint adjustment does not affect any of the conclusions
in this paper---the intrinsic widths of the color selection regions are
$\ga0.1$~mag (Fig.~\ref{fig4}) and all magnitude measurements are assumed to
have an associated systematic error of $\pm0.1$~mag in the data analysis
(Sec.~2.5).

We have carried out tests using three different photometry techniques:
{\smcap focas}, aperture photometry with curve of growth corrections, and
PSF-fitting using the digital stellar photometry program {\smcap daophot}
(Stetson 1987, 1992).  The results obtained from these techniques are not
vastly different.  The {\smcap focas} limiting isophotes are better matched
to the light distribution of galaxies of widely varying shapes and sizes
than fixed apertures and suffer less from contamination in crowded fields.
The {\smcap focas} ``common optimized aperture'' color measurements are
somewhat more robust than the {\smcap daophot} measurements for the resolved
galaxies in the data set; the two techniques yield comparable results for
unresolved objects.

The Galactic globular cluster images are analyzed with the help of the
{\smcap daophot} applying the {\smcap find} and {\smcap allstar} tasks in the
usual way.  We use a linearly variable point spread function (PSF) template
to account for slight distortion in image quality across the image.
Position-dependent aperture corrections are applied to convert the apparent
magnitudes returned by {\smcap allstar} to total magnitudes.

The {\smcap focas}-based (M31 halo and comparison fields) and {\smcap
daophot}-based (Galactic globular clusters) total instrumental magnitudes
are transformed to the standard Johnson/Kron/Cousins $UBRI$ system as
defined by the set of Landolt (1983) photometric standards.  The
transformation parameters in each band, derived from observations of standard
star fields, include a zeropoint accurate to $\la0.05$~mag, a shutter time
constant of~0.07~s (same for all bands), an independent airmass coefficient
for each of the two~photometric nights, and a linear (Johnson/Kron/Cousins)
color term.  The color term, particularly large for the $B_J^{\rm
instr}\rightarrow{B}^{\rm Johnson}$ conversion, is applied iteratively
starting with the instrumental color as an initial guess.  We
hereafter use the term ``$UBRI$ magnitudes'' to refer to calibrated
magnitudes on the Johnson/Kron/Cousins system (Landolt 1983). 

\subsection{Photometric Accuracy and Completeness}

We estimate the photometric error associated with the measurement of $UBRI$
magnitudes of stars/galaxies in the M31 halo and comparison fields.  In the
M31 halo field, the Poisson error is 0.05~mag for objects with $U=23.0$,
$B=24.1$, $R=23.3$, and $I=22.6$, and increases to 0.25~mag for objects with
$U=25.5$, $B=26.7$, $R=25.9$, and $I=25.1$.  These estimates are based on
shot noise in the photons detected from the object, and on the area of the
limiting {\smcap focas} isophote and the shot noise per pixel in the sky
counts.  The overall error in the final calibrated magnitude is likely to be
somewhat larger than the Poisson error because of the following:
flat-fielding error, residual image artifacts, error in measuring the value
of the local sky background, neighbor contamination (varying degrees of
contamination among the 4~bands can lead to color errors), unaccounted for
light outside limiting isophote (this does not affect color measurements to
first order), and calibration error.  An empirical check of the measured
$UBRI$ colors of Galactic dwarf stars in the M31 halo and comparison fields
shows that they agree with the Bertelli et~al.\ (1994) model predictions to
within $\sim\pm0.05$~mag; nevertheless, we conservatively assume that these
other sources of error contribute an additional (systematic) uncertainty of
0.1~mag (over Poisson error) in all magnitude measurements (see Sec.~3.2).

Based on the sky noise level in the final M31 halo field images, we compute
$3\sigma$ limiting apparent magnitudes within an aperture of area 28.3~pixels
or 6.25~arcsec$^2$ (this is roughly equal to the minimum allowable isophotal
area in the {\smcap focas} object detection algorithm): $U_{\rm lim}=25.6$,
$B_{\rm lim}=26.7$, $R_{\rm lim}=25.7$, and $I_{\rm lim}=25.0$.  Note, these
limiting magnitudes are based on a {\it fixed\/} aperture size, whereas the
error estimates in the previous paragraph are based on the {\it actual\/}
isophotal area of each object.  The remainder of the data analysis presented
in this paper is restricted to the sample of stars/galaxies that satisfy the
criteria: $B<B_{\rm lim}$, $R<R_{\rm lim}$, and $I<I_{\rm lim}$.  As
discussed in Sec.~3.2 below, the $U$ band is the most critical of the 4~bands
in discriminating between faint galaxies and red giants, even in cases where
there is only an upper limit to the apparent $U$ brightness of an object.  We
have therefore chosen {\it not\/} to impose the $U$ limiting magnitude
criterion in defining the sample.  Instead, any object whose measured $U$
brightness falls formally below the threshold is treated as a non-detection
in the $U$ band, and is assigned a well-defined upper limit to its
brightness, $U>U_{\rm lim}$, for the purpose of color selection (Sec.~3.2).

Even after applying the above cuts in apparent $BRI$ magnitude, the sample of
stars and galaxies detected in the M31 halo and comparison fields is not
complete over all regions of the CCD image: {\smcap focas}'s object detection
routine has difficulty finding objects in areas of the $U+B+R+I$ image
surrounding very bright galaxies and stars, especially those with charge
bleed artifacts.  The nominal area covered by the M31 halo data set (field of
view of KPNO image) is about 227~arcmin$^2$ but a fraction of this area,
$(1-P_0)=0.08$, is effectively lost.  Even though $P_0$ is probably a weak
function of apparent magnitude, it is adequate for our purposes to assume it
is a constant.  We estimate $P_0$ by comparing the total number of objects
detected down to the completeness limits over the entire KPNO image to the
{\it expected\/} total number based on the observed surface density of
objects over the same magnitude range in ``clean'' areas of the image
(i.e.,~areas free of very bright stars/galaxies, charge bleeds, and
hysteresis artifacts).

Since the majority of faint objects in the M31 halo field are distant field
galaxies (an isotropic population), we can estimate the degree of
completeness of the sample by comparing the observed counts, $N(m)$, to
counts of a statistically complete sample of galaxies derived from the
high-quality, Hubble Deep Field (HDF) data set (Williams et~al.\ 1996).  For
the purpose of this comparison only, the above limiting magnitude criteria
are ignored and the entire KPNO data set is used.  The {\it HST\/}
instrumental magnitudes of HDF galaxies are converted to the
Johnson/Kron/Cousins $UBRI$ magnitude system adopted in this paper using the
transformation relations given by Holtzman et~al.\ (1995); the observed
distribution of galaxies in the $B-R$ vs $B-I$ plane (in both M31 halo and
comparison fields) is used to interpolate each HDF galaxy's $R$ band
magnitude.  For the 4~bands, the apparent magnitude at which the completeness
drops to half the maximum value is: $U_{50}=26.3$, $B_{50}=27.6$,
$R_{50}=26.5$, and $I_{50}=25.8$.  These values are significantly
fainter than the $3\sigma$ limiting apparent magnitudes at which the
data set is truncated; the completeness fraction is essentially $100\%$
for objects brighter than the limiting magnitudes.

\subsection{Angular Size Measurements}

The angular diameter ($\theta_{\rm FWHM}$) of each object detected in the M31
halo field and comparison field is measured using the final, coadded KPNO
$I$ band images, as they have the best seeing of the 4~bands (FWHM of seeing
is about $1\farcs2$ and $1\farcs5$ in the two fields, respectively).  The
seeing in a few of the individual 10~min KPNO $I$ exposures is as good as
$0\farcs8$ (FWHM), but these images do not have adequate signal-to-noise for
the majority of the objects which are quite faint.  For each object,
$\theta_{\rm FWHM}$ is determined by fitting a Gaussian to the
azimuthally-averaged radial intensity profile.  The outer parts of the
stellar PSF are not very well described by a Gaussian so the $\theta_{\rm
FWHM}$ values reported in this paper may be systematically different from
measurements based on a more realistic PSF template; this study however
relies only on the {\it relative\/} angular sizes of objects (Sec.~3.3).

We also determine widths of the best fit Gaussian profiles for objects that
lie in the area of the M31 halo field covered by the Keck/LRIS $I$ band
images.  The Keck data set has better angular resolution
($0\farcs6\>$--$\>0\farcs9$ seeing FWHM and $0\farcs22$~pixel$^{-1}$ scale)
than the coadded KPNO image and is of comparable or greater depth (the factor
of~6 larger primary mirror area and better seeing more than compensate for
the much shorter exposure times).  We use Keck-based $\theta_{\rm FWHM}$
measurements in preference to those derived from the KPNO data whenever
available; the former allow improved morphological distinction between stars
and compact galaxies at faint apparent magnitudes.

\section{Searching for M31 Halo Red Giants: Needles in a Haystack}
\subsection{Defining the Stellar Locus in UBRI space}

Accurate definition of the loci of RGB stars in $UBRI$ space is very
important for the color selection technique used in this paper.  Following
the method used by GGRF who determined stellar loci using the Yale stellar
evolution models (Green et~al.\ 1987), we determine RGB stellar loci using
the Bertelli et~al.\ (1994) stellar evolution models.  The Bertelli et~al.\
model isochrones match the color distribution of Landolt (1983) standard
stars extremely well, in the Johnson/Kron/Cousins $UBVRI$ photometric system
defined by these stars.  Since the photometric calibration of the KPNO
data set used in this paper is tied to these same Landolt standards (in terms
of magnitude zeropoints and color transformations), the Bertelli et~al.\
models are a natural choice for defining the stellar loci used in this study.

\begin{figure}
\epsscale{0.8}
\plotfiddle{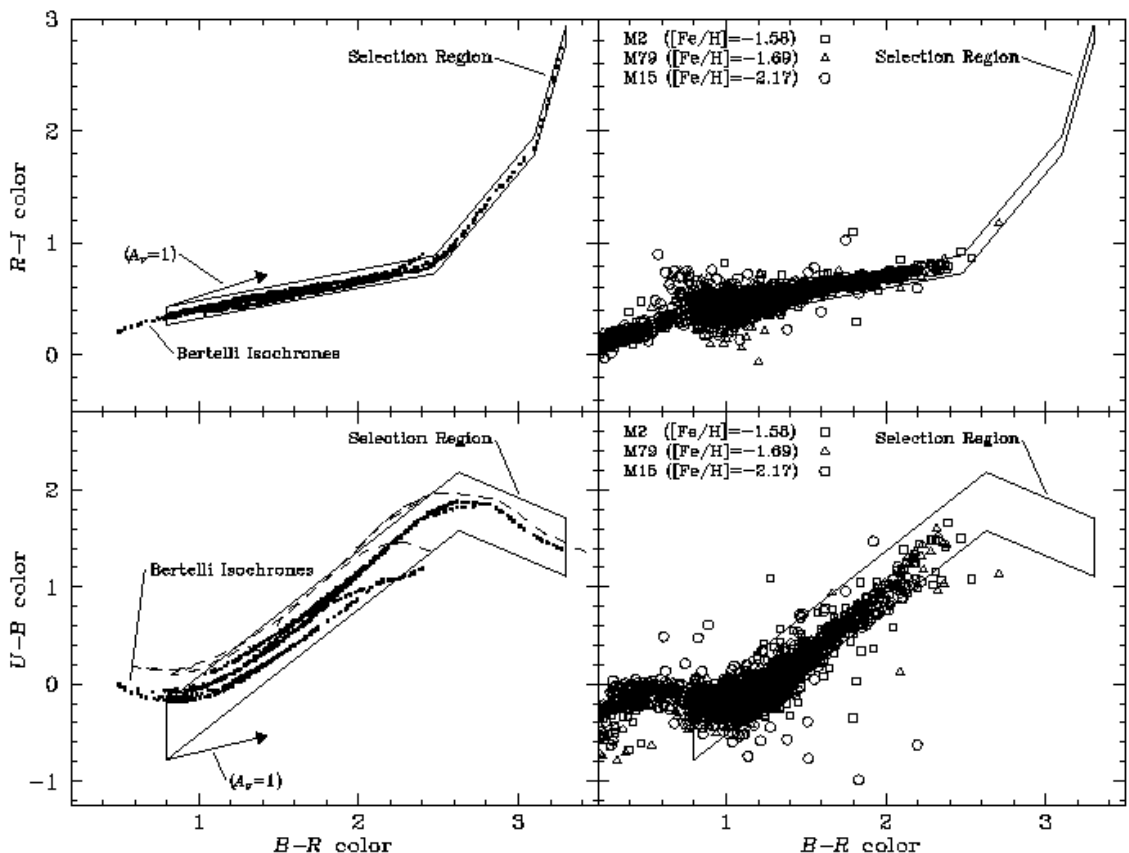}{4in}{-90}{70}{70}{-290}{385} 
\vskip 0.2in
\caption{The loci of stars in the $BRI$ and $UBR$
color-color diagrams.  [All colors are adjusted to the amount of
reddening appropriate for the M31 halo field, $E(B-V)=0.085$.]~~ {\it
Left panels}---Theoretical isochrones from Bertelli et~al.\ (1994) for
ages of 5$\>$--$\>$20~Gyr and for metallicities of $\rm[Fe/H]=-1.70$
to $\rm[Fe/H]=0$ (bold dots).  The dashed line in the lower left panel
corresponds to a 1~Gyr stellar population with super-solar metallicity,
$\rm[Fe/H]=+0.4$.  The isochrones exclude main sequence
stars with $M\la0.6\,M_\odot$ causing an artificial truncation
(e.g.,~at $B-R=2.4$, 
isochrones $U-B=1.2$).  The solid lines define the ``selection
parallelogram'', the region used to select RGB stars in the M31 halo
and comparison fields.  The reddening vector is shown for $A_V=1$~mag
(its location in the color-color plane is arbitrary).~~ {\it Right
panels}---Distribution of relatively metal poor stars in the Galactic
globular clusters, M2 (squares), M79 (triangles), and M15 (circles),
along with the RGB selection regions shown in the left panels.  The
distribution of cluster RGB stars is in good agreement with
theoretical models; the feature extending beyond the blue end of the
selection region ($B-R<0.8$) is formed by cluster horizontal branch
stars.  \label{fig2}}
\end{figure}

We use Bertelli et~al.\ model isochrones bracketing the 5$\>$--$\>$20~Gyr
range of ages and for the metallicity range $\rm[Fe/H]=-1.70$ to
$\rm[Fe/H]=0$ (bold dots in left panels of Fig.~\ref{fig2}) to define the
stellar selection loci in the $R-I$ vs $B-R$ and $U-B$ vs $B-R$ color-color
planes.  The isochrones are reddened by an amount appropriate for the M31
halo field, $E(B-V)=0.085$ (Burstein \& Heiles 1982), using a standard
Galactic interstellar dust extinction law (Cardelli et~al.\ 1989).  The
selection region (the narrow region in Fig.~\ref{fig2} with
parallel sides which we refer to as a ``parallelogram'' for simplicity) is
defined as the region that completely encompasses all of the RGB points in
the isochrones.  While we do not expect there to be many young or high
metallicity objects, we have chosen the selection region to avoid biasing the
resulting sample to any particular metallicity or age.

Even though the stellar locus is very extended in any one color (e.g.,~it
spans the range $0.8<B-R<3.3$), the different colors are strongly
correlated and this makes the locus narrow in color-color space.
Changes in metallicity and age tend to move stars mostly along the length of
the stellar loci: this is especially true in the $BRI$ color-color diagram;
increasing [Fe/H] does make the $U-B$ color slightly redder at fixed $B-R$,
so the width of the $UBR$ color-color selection region is correspondingly
wider to account for this.

We perform two independent empirical checks to verify the accuracy of the
stellar locus definitions and the photometric conversion of instrumental
magnitudes to the Johnson/Kron/Cousins system:
\begin{itemize}
\item[(1)]{Globular cluster data, obtained in the same $UB_JRI$ system in
conjunction with the M31 halo field and comparison field observations, yield
a distribution of RGB stars that is in good agreement with the Bertelli
et~al.\ (1994) model isochrones.  This is shown in the right panels of
Fig.~\ref{fig2} (the sequence of stars beyond the left edge of
the selection box corresponds to the cluster horizontal branch).  Prior to
the comparison, the colors of the globular cluster stars are dereddened and
then adjusted to the line-of-sight reddening estimated for the M31 halo
field, $E(B-V)=0.085$ (Burstein \& Heiles 1982).  Note however that data are
only available for the relatively metal-poor clusters, M15, M2, and M79,
thereby restricting the check to the range: $\rm-2.5\la[Fe/H]\la-1.5$.}

\item[(2)]{As discussed in Sec.~3.2 below, the distribution of foreground
Galactic metal-rich ($\rm[Fe/H]\sim0$) disk dwarf stars in $UBRI$ color-color
diagrams of the M31 halo and comparison fields is consistent with Bertelli
et~al.\ isochrones of solar or super-solar metallicity (compare dashed line
in lower left panel of Fig.~\ref{fig2} to lower left panel of
Fig.~\ref{fig3}).  This check is indirect however because the foreground
stars are (mostly) main sequence stars whereas the selection loci are
optimized for red giants.}
\end{itemize}

\subsection{Color Selection}

The main source of contamination for the study of M31's faint spheroid stars
is the population of background faint field galaxies.  These galaxies are
expected to be about as common as M31 halo stars at faint magnitudes (there
are 25~arcmin$^{-2}$ down to $B=26$) and a fair fraction have small angular
sizes, which makes it difficult to distinguish them from stars on the basis
of image morphology alone.  The expected surface density of M31 spheroid red
giants in this 19~kpc field is less than 26~arcmin$^{-2}$ down to $I=24$
($B\sim26$) as discussed in Sec.~4.2.  GGRF have developed a method which
uses accurate $UBRI$ CCD photometry for isolating faint stars ($B<26$) and
rejects most of the background galaxies on the basis of colors while
eliminating only a fraction of the stars.  The essence of the method lies in
the fact that stars occupy a relatively tight locus in any color-color
diagram (Fig.~\ref{fig2}), while galaxies exhibit a broad
distribution of colors.

\begin{figure}
\epsscale{0.8}
\plotfiddle{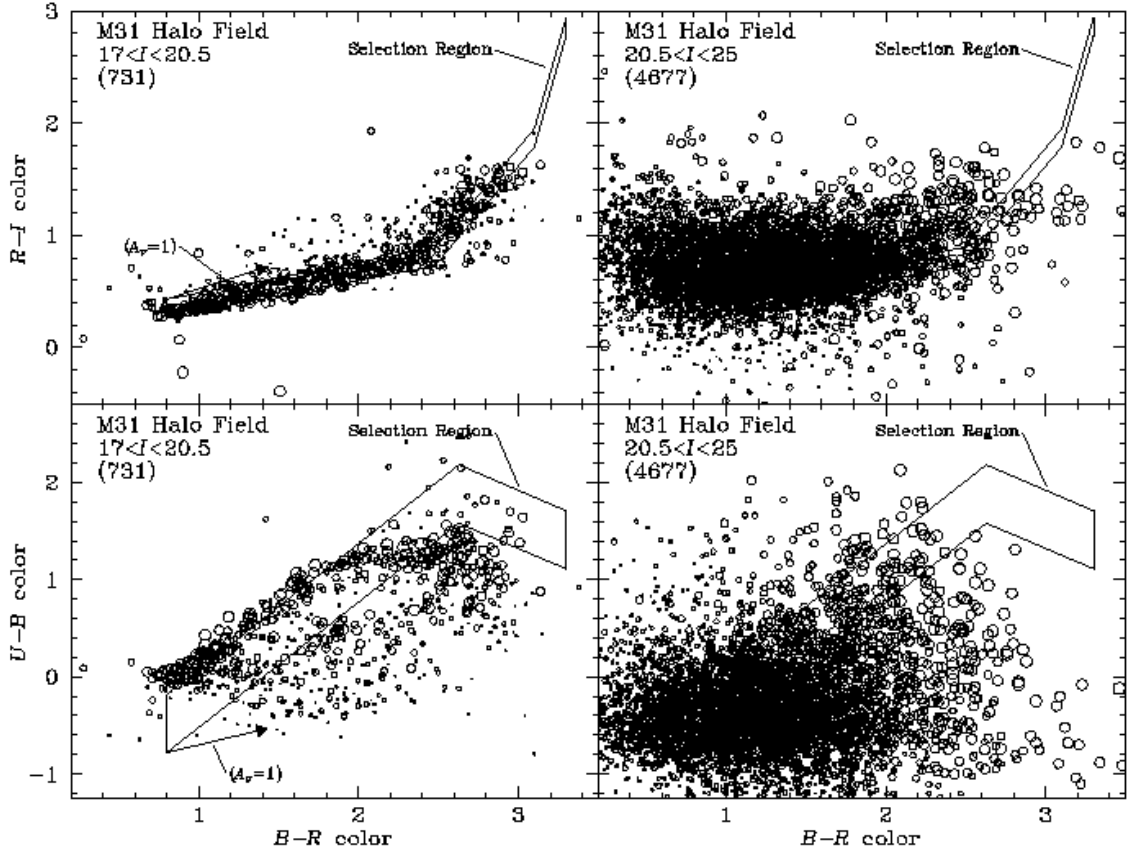}{4in}{-90}{70}{70}{-290}{385}
\vskip 0.2in
\caption{Color-color diagrams for the M31 halo
field: $R-I$ vs $B-R$ (upper panels) and $U-B$ vs $B-R$ (lower
panels).  Point size is scaled with apparent $I$ magnitude, with
larger dots indicating brighter objects.  The left panels show objects
with $17<I<20.5$.  These bright objects are mostly foreground main
sequence stars in the Galactic disk; their distribution matches the
selection region (see dashed line in Fig.~\ref{fig2}), though not
perfectly as selection is optimized for
lower metallicity RGB stars.  The right panels show objects with
$20.5<I<25$.  The majority of the objects in this faint magnitude
range are galaxies which have a much larger spread in color than the
stars.  Note, most galaxies with $B-R\ga2$ are $\sim1$~mag bluer in
$U-B$ than the selection region. \label{fig3}
}
\end{figure}

\begin{figure}
\epsscale{0.8}
\plotfiddle{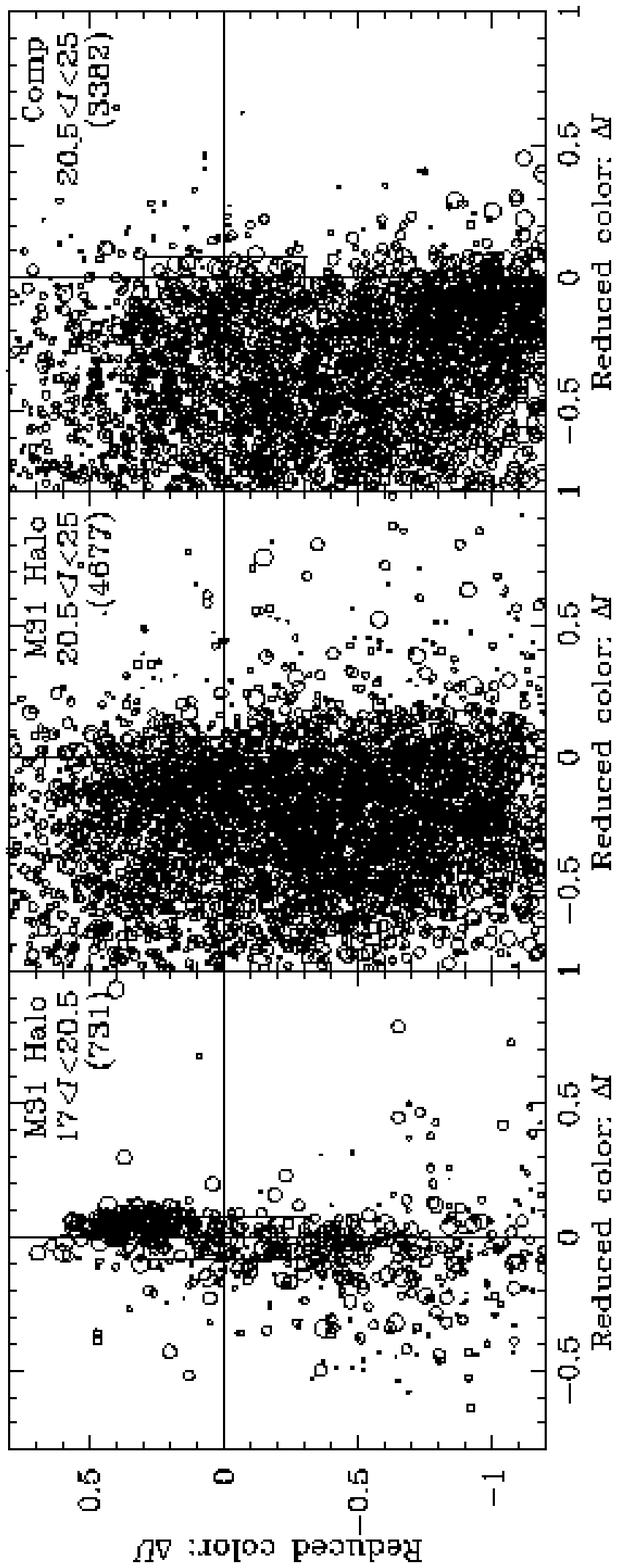}{2.5in}{-90}{70}{70}{-290}{250}  
\caption{ The distribution of color excess
indices, $\Delta{I}$ and $\Delta{U}$.  Dot size scales with apparent
$I$ magnitude: large dots~=~bright objects, and vice versa.  The left
panel shows objects with $17<I<20.5$ in the M31 halo field: most of
these are Galactic dwarfs with high metallicities, with colors that
deviate slightly from the expected locus of RGB colors.  The center
and right panels show objects with $20.5<I<25$ in the M31 halo field
and the comparison field, respectively; the majority of these are
faint galaxies (especially in the comparison field) exhibiting a broad
distribution of color excesses.  The width and height of the bold
rectangle at the origin correspond to the vertical widths of the
selection regions used in the $BRI$ and $UBR$ planes, respectively.
 \label{fig4}}
\end{figure}

Figure~\ref{fig3} demonstrates the basis for the color selection
method via $BRI$ and $UBR$ color-color plots of all objects detected in the
M31 halo field.  For an object to appear in these color-color plots and in
the $\Delta{U}$ vs $\Delta{I}$ color excess plots below
(Fig.~\ref{fig4}) it must satisfy the $U<U_{\rm lim}$ criterion
along with the usual $BRI$ criteria; this additional restriction causes there
to be substantially fewer objects in these plots than in the full sample
(cf.~Figs.~\ref{fig6} and \ref{fig7}), especially at
the faint end of the distribution.  The left panels include all objects
brighter than the expected tip of the M31 RGB, $I<20.5$.  Most of the bright
objects in this low latitude ($\vert{b}\vert=22^\circ$) M31 halo field are
foreground Galactic stars as evidenced from the fact that they lie close to
the defined selection region.  The selection regions are, however, optimized
for red giant stars of solar or lower metallicity, while most of these bright
objects are disk dwarfs with high (possibly super-solar) metallicity.  Thus
it is not surprising that the $I<20.5$ objects deviate from the selection
region slightly, especially in the $UBR$ diagram.  The right panels of
Fig.~\ref{fig3} show all objects with $I>20.5$, which includes any
M31 RGB stars as well as a large number of distant field galaxies.  It is
obvious that the galaxies occupy a more extensive region of the plot than do
the stars.  In the $BRI$ diagram, the galaxies are slightly redder in $R-I$
($\sim0.4$~mag) than the stellar locus at a given $B-R$ color; the
separation is clearer in the $UBR$ diagram: for example, the typical galaxy is
$\ga1$~mag bluer in $U-B$ than the stellar locus for $B-R\ga2$.

Following GGRF we use each object's measured $B-R$ color and the stellar locus
to predict the colors, $(R-I)_{\rm pred}$ and $(U-B)_{\rm pred}$, an RGB star
would have if it had the same $B-R$ color as the object.  The measured
colors of the object are then used to calculate color excesses:

\begin{equation}
\Delta I = -[(R-I) - (R-I)_{\rm pred}]
\end{equation}
\begin{equation}
\Delta U = (U-B) - (U-B)_{\rm pred}
\end{equation}

\noindent 
The color excesses measure the vertical distance in the $BRI$ and $UBR$
color-color diagrams of each object from center line of the stellar locus
``parallelogram'' (Fig.~\ref{fig2}).  The $U$ data are pivotal
in the application of the color selection technique to red giants, even
though the low intrinsic stellar brightness and relative inefficiency of the
instrument in the ultraviolet conspire to make $U$ band detection of distant
RGB stars a challenging prospect.  As described in Sec.~2.5, objects that are
fainter than the limiting magnitude in $U$ (but which satisfy the apparent
brightness criteria in $BRI$) are assigned a $U$ band color excess
$\Delta{U}>\Delta{U}_{\rm lim}$, where $\Delta{U}_{\rm lim}=(U_{\rm
lim}-B)-(U-B)_{\rm pred}$.

Figure~\ref{fig4} shows the distribution of color excesses.  The
relatively bright objects ($17<I<20.5$) in the M31 halo field shown in the
left panel lie near but not exactly at the origin.  This is only to be
expected: while most of these bright objects are foreground  Galactic stars,
they are main sequence stars and not red giants.  In the center and right
panels, we plot the fainter objects ($20.5<I<25$) in the M31 halo and
comparison fields, respectively.  Note the broad and asymmetric distribution
of these faint objects in ($\Delta{I}$,~$\Delta{U}$) space most of which
belong to the faint background field galaxy population (especially in the
comparison field, right panel).

Color selection is accomplished by requiring objects to have colors
consistent with both $BRI$ and $UBR$ stellar loci.  An object is considered
to pass color selection only if its color excesses ($\Delta{I}$,~$\Delta{U}$)
are such that the $1.5\>\sigma$ measurement error ellipse intersects a
rectangular reference region centered on the origin
(Fig.~\ref{fig4}), where the width and height of the rectangle are
equal to the vertical width of the ``parallelogram'' selection regions in the
$BRI$ and $UBR$ color-color plots, respectively
(Fig.~\ref{fig2}).  In other words, an object is assigned a
color-based classification of ``star-like'' only if $\delta<1.5$, where
$\delta$ is the {\it significance}, in units of the photometric error, of the
separation of the object from the point closest to it within the central
rectangular area in ($\Delta{I}$,~$\Delta{U}$) space.  Like GGRF, we
determine the measurement errors for $\Delta{I}$ and $\Delta{U}$ jointly, as
there is a strong covariance between these quantities.  For each $U$ band
non-detection that is brighter than the limiting magnitudes in $BRI$, we
check whether the combination of its $\Delta{I}$ value and
$\Delta{U}>\Delta{U}_{\rm lim}$ criterion is consistent with the rectangle at
the origin.

\subsection{Morphological Selection}

The $UBRI$ color selection described above yields a sample of objects
consisting of M31 RGB stars, foreground Galactic stars (mostly disk dwarfs),
and any field galaxies whose colors happen to be consistent with stellar
loci.  We also apply the more traditional method of star-galaxy separation
based on image morphology.

\begin{figure}
\epsscale{0.8}
\plotone{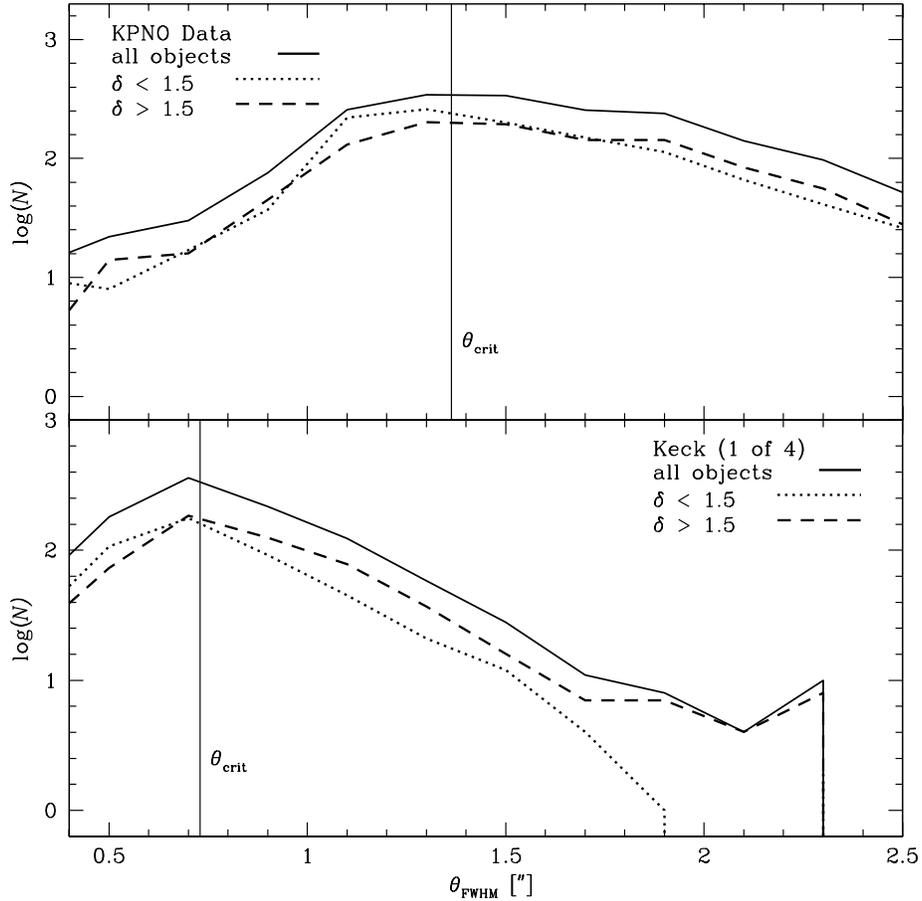}
\caption{The distribution of angular sizes for
objects detected in the M31 halo field as measured on the KPNO image
(upper panel) and on one of the four~Keck images (lower panel).  The
threshold for morphological selection, $\theta_{\rm crit}$, is
indicated by a vertical line.  The solid curve shows the entire
(unselected) sample; the dotted curve is for objects whose colors are
consistent with those of an RGB star ($\delta<1.5$), while the dashed
curve is for the rest of the objects ($\delta>1.5$).  Note, objects
with $\delta<1.5$ tend to be more compact than those with
$\delta>1.5$, an indication that the color selection does
preferentially select stars. \label{fig5}}
\end{figure}

Figure~\ref{fig5} shows the distribution of angular sizes ($\theta_{\rm
FWHM}$) of objects in the M31 halo field as measured on the KPNO image and on
one of the Keck/LRIS $I$ band images.  About half the objects in the sample
are well resolved relative to the stellar PSF, $\theta_{\rm FWHM}>\theta_{\rm
crit}$ (morphologically ``galaxy-like''); the rest are assigned a
morphological classification of ``star-like''.  We adopt $\theta_{\rm
crit}=1\farcs36$ for the KPNO image, and $\theta_{\rm crit}=0\farcs73$,
$0\farcs84$, $0\farcs86$, and $0\farcs88$ for the 4~Keck/LRIS images.  These
$\theta_{\rm crit}$ values are slightly larger than the FWHM of the PSF in
the center of the corresponding CCD image to allow for variations in the PSF
across the field of view: the stellar image quality is degraded near the
corners of the field due to optical distortion/focus variation.  For an
object larger than a few arcseconds, the measured size is typically
inaccurate since it is based on a Gaussian fit.  This does not affect our
analysis, however, since $\theta_{\rm FWHM}$ for such objects is invariably
large; hence the objects are correctly classified as morphologically
galaxy-like.

Keck/LRIS-based star-galaxy separation is used in preference to KPNO-based
star-galaxy separation in areas where both measurements are available (75\%
of the M31 halo field); the superior angular resolution and signal-to-noise
ratio of the Keck $I$-band data result in improved star-galaxy
discrimination, especially in the case of compact, barely resolved galaxies
and blended groups of faint objects.  It is possible to compare star-galaxy
discrimination based on Keck vs KPNO $\theta_{\rm FWHM}$ measurements in the
area where the two data sets overlap: the morphological classification is
star-like by both methods for 34\% of the objects, galaxy-like by both
methods for 26\%, star-like according to the Keck measurement and galaxy-like
according to the KPNO measurement for 27\%, and vice versa for 13\%.

The distribution of measured angular sizes may be used to investigate whether
the color selection technique is working.  In the unselected sample, 56.0\%
of all objects with $I\leq20$ have $\theta_{\rm FWHM}$ values that lead to
their being classified as morphologically star-like, whereas after color
selection $61.6\%$ of the objects are morphologically star-like.  For
$I\leq23$, only 44.4\% of all objects have stellar profiles, yet 53.4\% of
the color selected objects are star-like.  For $I\leq25$, 48.8\% of the full
(non-color selected) sample has star-like morphology, compared to 55.3\% for
the color selected sample.  The fact that the fraction of morphologically
star-like objects is higher in the color selected sample than in the full
sample is indicative that color selection is indeed picking out stars.  The
effect is subtle since neither morphological selection nor color selection
alone is perfectly efficient.  A fraction of the stars ($\sim30$\%) happen to
be scattered beyond $\delta>1.5$ due to photometric error, and there are
compact galaxies that fail the $UBRI$ selection criterion as well as extended
galaxies whose colors happen to intersect the RGB locus.  Furthermore, errors
in the measurement of angular size (due to object/sky photon noise, neighbor
contamination, etc.) cause $\theta_{\rm FWHM}$ for some stars to scatter
beyond $\theta_{\rm crit}$, especially at faint magnitudes.

There is nothing special about our use of $\theta_{\rm FWHM}$ measurements
for star-galaxy discrimination nor about our particular choices of
$\theta_{\rm crit}$ values.  After carrying out tests of {\smcap focas}'s
Bayesian star-galaxy separation scheme on the data sets, we choose to adopt
the simpler $\theta_{\rm FWHM}$ measurement scheme; the effect of noise on
star-galaxy discrimination is easier to quantify in the latter scheme.  It is
worth noting however that results based on the {\smcap focas} morphological
classification are not qualitatively different from those based on
$\theta_{\rm FWHM}$ measurements.  The $\theta_{\rm FWHM}$-selected sample no
doubt includes some compact galaxies for which the measured angular size is
less than the critical value.  Conversely, a small fraction of stars are
scattered across to $\theta_{\rm FWHM}>\theta_{\rm crit}$ due to measurement
error in the angular size caused by noise and/or the presence of neighbors.
We have chosen $\theta_{\rm crit}$ values to achieve a balance between
contamination vs incompleteness.

\subsection{Detection of Red Giants in M31}

\begin{figure}
\epsscale{0.8}
\plotone{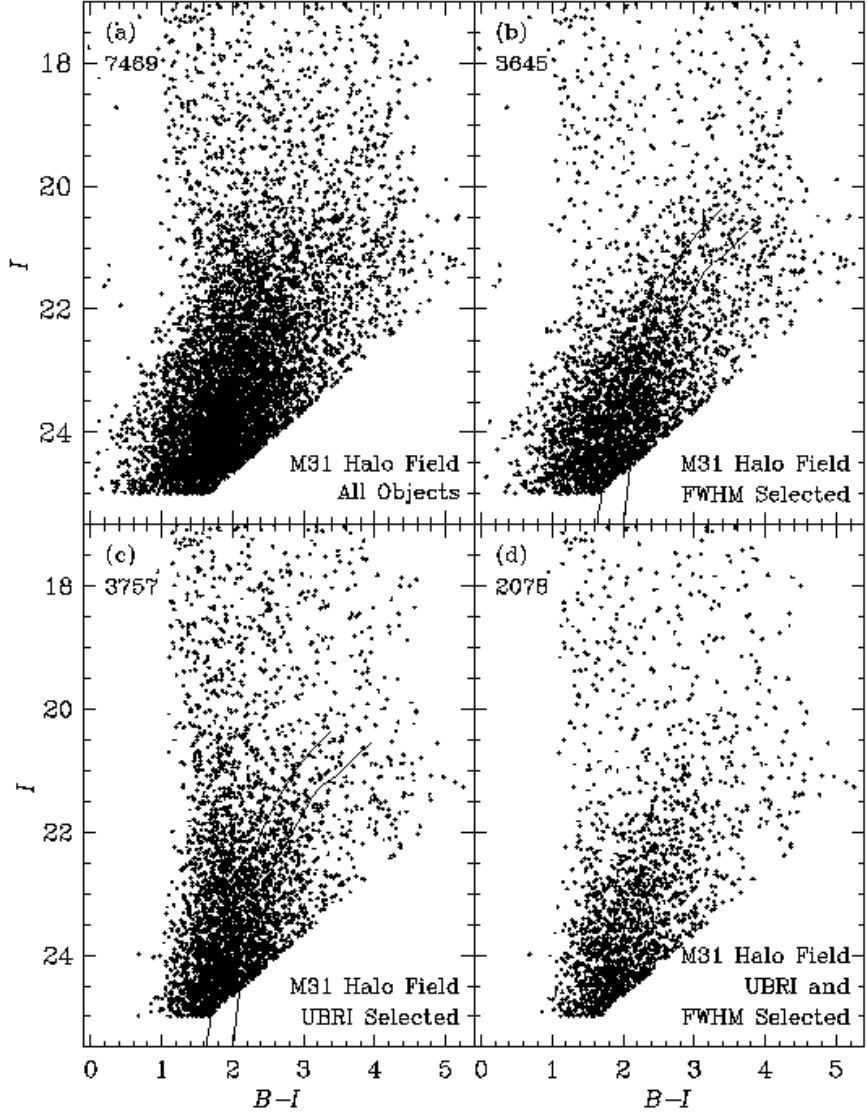}
\caption{(a)~Color-magnitude diagram of all M31
halo field objects.~~ (b)~CMD after morphological selection.~~ (c)~CMD
after $UBRI$ color selection.~~ (d)~CMD after $UBRI$ and FWHM
selection.~~ The number of objects in each panel is indicated in the
upper left.  Most of the bright objects ($I\la20$) are foreground
Galactic stars, the prominent vertical edge at $B-I=1$ representing
the main sequence turnoff, while the majority of faint objects are
distant galaxies.  The diagonal concentration of faint red objects
that survive in~(d) resembles the RGB fiducials of Galactic globular
clusters displaced to $D_{\rm M31}=770$~kpc shown in~(b) and (c):
[L$\rightarrow$R] M5 ($\rm[Fe/H]=-1.4$) and 47~Tuc
($\rm[Fe/H]=-0.70$). \label{fig6}}
\end{figure}

The detection of M31 RGB stars against the numerous background field galaxies
and foreground Galactic stars is difficult, but the combined power of color
and FWHM selection yields a sample of M31 halo RGB stars.  The pre-selection
color-magnitude diagram (CMD) of all objects detected in the M31 halo field
is shown in Fig.~\ref{fig6}(a); it is practically impossible to see any
concentration of M31 RGB stars in this plot.  After FWHM selection a hint of
concentration of M31 RGB stars becomes visible in the CMD
[Fig.~\ref{fig6}(b)].  Morphological selection removes 51.2\% of the
objects in the full sample, most of them relatively well resolved galaxies,
yet contamination from compact galaxies and foreground main sequence stars
remains.  The $UBRI$ selection technique removes 49.7\% of the full set of
objects [Fig.~\ref{fig6}(c)].  Most foreground lower main sequence
stars with $B-I>3$ are removed (these deviate systematically from the RGB
loci) as are a significant fraction of the distant field galaxies (these have
a broad distribution in color-color space), yet galaxies whose $UBRI$ colors
happen to be close to those of a RGB star survive selection.  While a
concentration of M31 RGB stars in the CMD is again revealed, as for the FWHM
selected sample, the exact location of the concentration is uncertain due to
contamination.  When FWHM and color selection are applied simultaneously, the
complimentary nature of these selection screens reveals a fairly clean sample
of M31 RGB stars [Fig.~\ref{fig6}(d)].  Comparing the Galactic
globular cluster RGB fiducials of M5 and 47~Tuc [plotted in
Fig.~\ref{fig6}(b) and (c), but not in (d) so as not to guide the
reader's eye] to the main concentration in the FWHM- and $UBRI$-selected CMD
[Fig.~\ref{fig6}(d)], it is clear that the objects detected are in a
location of the CMD appropriate for M31 RGB stars. 

We carry out a detailed comparison between the M31 halo field sample and the
sample of objects found in the comparison field.  The final image of the
comparison field has an area of 254~arcmin$^2$; its effective area is 94.1\%
of this, or 239~arcmin$^2$ ($P_0$ has been estimated in the same way as for
the M31 halo field---see Sec.~2.5).  The M31 halo field has a smaller
effective area (88\%) than the comparison field.  The latter data set is
scaled down in order to match the effective area of the M31 halo field by
removing a random 12\% of the objects from the sample.  This reduces the
number of objects in the full (unselected) comparison field sample to 4217.
The matched comparison field data set is then processed through the same
$UBRI$ color selection and FWHM selection screens as the M31 halo field data
set.  The comparison field data were obtained under slightly worse seeing
conditions than the M31 halo field data; the KPNO $I$ band comparison field
data set is used for morphological selection with $\theta_{\rm
crit}=1\farcs65$.  No Keck/LRIS images of the comparison field are available
for FWHM measurements.

\begin{figure}
\epsscale{0.7}
\plotfiddle{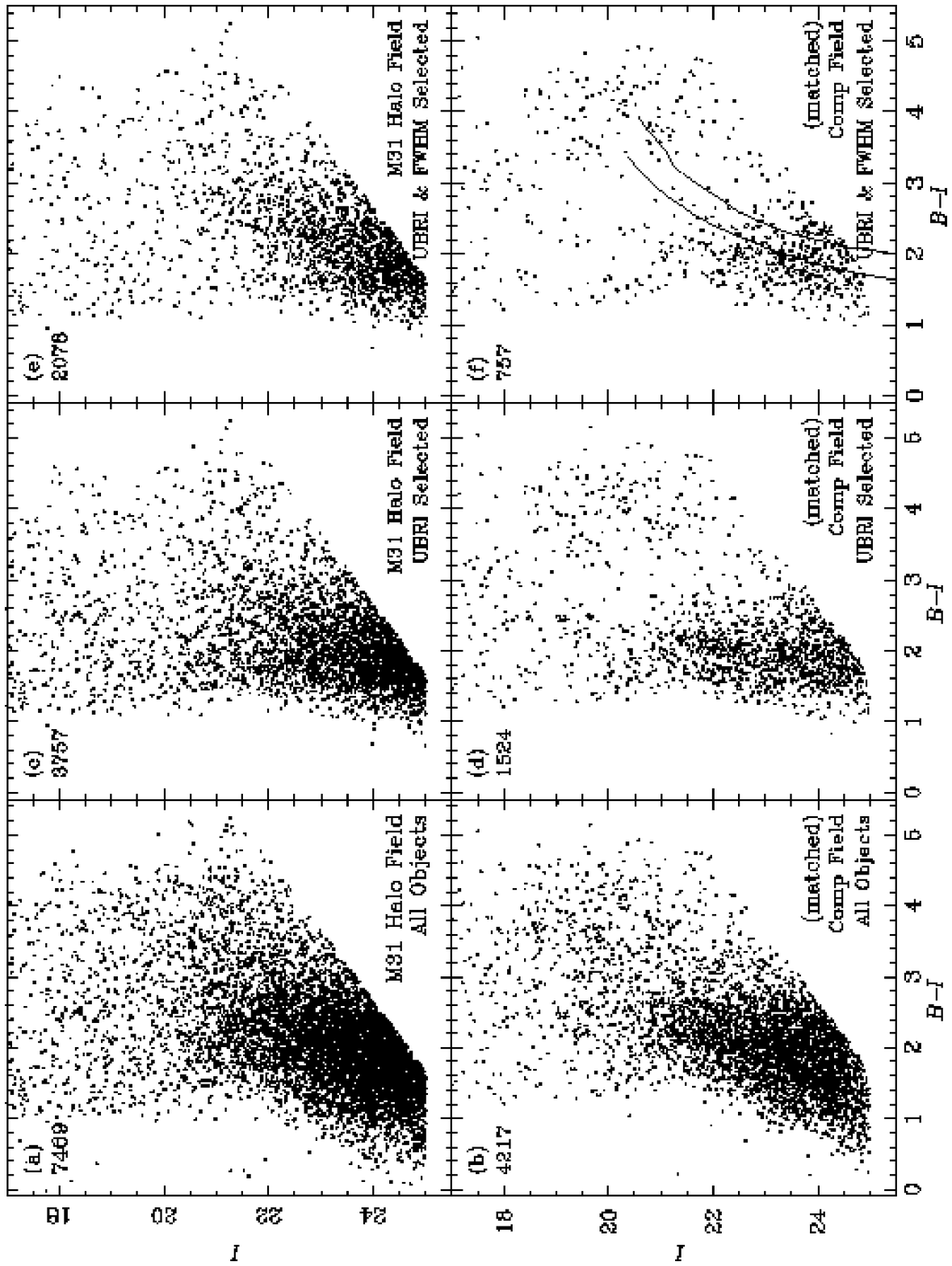}{4in}{-90}{70}{70}{-280}{410}
\caption{(a+b)~Color-magnitude diagrams of all
objects in the M31 halo and comparison fields, respectively, with the
latter scaled down to correct for the difference in the effective
areas of the two data sets.~~ (c+d)~Same as (a+b) after $UBRI$
selection.~~ (e+f)~Same as (a+b) after $UBRI$ and morphological
selection.~~ The number of objects in each panel is indicated in the
upper left.  The unselected M31 halo field contains many more objects
than the corresponding comparison field sample but it is difficult to
identify the additional population in the former sample [compare (a)
vs (b)].  Note the significant concentration in~(e) of additional
objects (RGB stars) in the final $UBRI$- and FWHM-selected M31 halo
field data set.  This may be compared to the RGB fiducials of M5 and
47~Tuc plotted in~(f), as for Fig.~\ref{fig6}.  \label{fig7}}
\end{figure}

Figure~\ref{fig7} shows $I$ vs $B-I$ CMDs of the M31 halo and
comparison fields (upper and lower panels, respectively), clearly
demonstrating the presence of an additional component of red giants in the
former data set.  The left panels [(a) and (b)] contain the full data sets
prior to any selection.  Even in this non-selected sample, it is obvious that
the M31 halo field contains many more objects than the comparison field, 7469
versus 4217.  This is mostly due to the additional population of M31 RGB
stars, although as we discuss below, the M31 halo field does contain a larger
number of foreground Galactic stars since it is at a somewhat lower Galactic
latitude.  In the center panels [(c) and (d)] we compare the fields after
$UBRI$ selection: 50.3\% of the M31 halo field objects survive, while only
36.1\% of the comparison field objects survive, an indication of the larger
population of objects with star-like colors in the former field.  The
additional population of M31 RGB stars is plainly visible, even though many
galaxies with star-like colors remain.  The right panels [(e) and (f)]
contain only objects that have both star-like colors and morphology.  There
is a noticeable additional component of M31 halo stars in
Fig.~\ref{fig7}(e).  As much as 57.0\% of the FWHM selected M31
halo field objects survive the additional color selection, while only 41.1\%
survive in the comparison field.

The objects that survive color and morphological selection in the comparison
field [Fig.~\ref{fig7}(f)] represent a mix of Galactic stars
and compact galaxies that happen to have star-like colors.  RGB fiducials for
the Galactic globular clusters M5 and 47~Tuc are also plotted to indicate the
region of the CMD where M31 RGB stars are expected to be concentrated.  The
significant concentration of additional objects in the M31 halo field sample
[Fig.~\ref{fig7}(e)] is indeed at this expected CMD location;
the M31 halo field also contains the foreground and background populations
that are present in the comparison field CMD.  The $UBRI$- and FWHM-selected
M31 halo field sample contains 1321 more objects than the corresponding
comparison field sample.  The number of these additional M31 halo field
objects increases towards fainter apparent $I$ magnitudes (see Table~1).
There are 145~additional objects in the range $20<I<22$, and 574~additional
objects in the range $22<I<24$.  Even in the magnitude range brighter than
the expected tip of the M31 RGB, $I<20$, there are 116~more objects in the
M31 halo field than in the comparison field.  This is due to the higher
surface density of foreground Galactic stars in the M31 halo field.

The degree of foreground contamination due to Galactic stars in the M31 halo
and comparison fields is investigated using an empirical model of the Galaxy
(IASG model---Ratnatunga \& Bahcall 1985).  The model incorporates standard
density profile parameters and metallicity distributions for the Galaxy's
spheroid, thin disk, and thick disk to compute the predicted stellar color
and apparent magnitude distributions.  These model CMDs are compared to the
CMDs of the M31 halo and comparison fields.  While the IASG model has not
been tested at faint magnitudes, the bright end stellar distributions agree
well with the M31 halo and comparison field data
[Figs.~\ref{fig8}(a), (b), (c), and (d)].  Note the ``wall'' of
main sequence turnoff stars near the left edge of each panel.  The predicted
density of stars with $I<20$ and $B-I<3.5$ agrees with that observed in the
corresponding fields.  The IASG model seems to over predict the number of
stars with $B-I>3.5$; in fact, recent star count analyses by Gould et~al.\
(1996, 1997) indicate that the abundance of low luminosity red ($B-I>3.5$)
main sequence stars in the Galaxy is lower by a factor of $\sim3$ than that
predicted by the IASG model.  Another reason for the discrepancy between the
M31 halo CMD and the IASG model CMD [Figs.~\ref{fig8}(a) vs
(c)] is that many of these faint red stars are screened out of the M31 halo
sample during $UBRI$ selection because the locus of such stars departs from
the RGB locus on which the selection ``parallelogram'' is based (see left
panels of Figs.~\ref{fig2} and \ref{fig3}).

\subsection{Statistical Removal of Contaminants from M31 RGB Sample}

\begin{figure}
\epsscale{1.0}
\plotfiddle{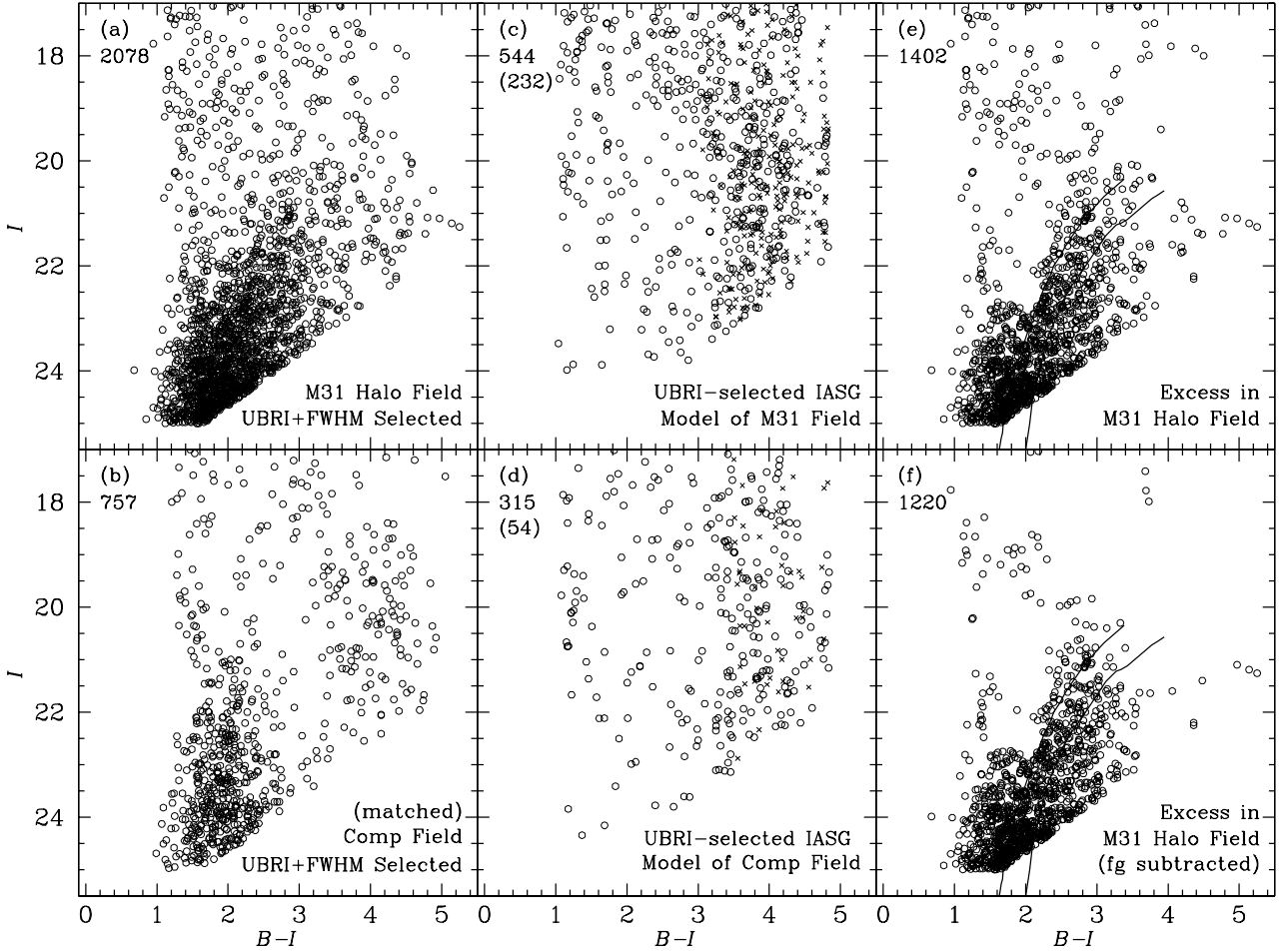}{4in}{-90}{70}{70}{-280}{420} 
\caption{Color-magnitude diagrams illustrating the
statistical removal of contaminants.  The number of objects in each
panel is indicated in the upper left.  RGB fiducials are the same as
in Fig.~\ref{fig6}.~~ (a+b)~Same as for Fig.~\ref{fig7}(e+f).~~
(c+d)~CMDs derived from IASG model predictions for the M31 halo and
comparison fields, respectively.  The faint crosses indicate the
fraction of faint red lower main sequence stars that fail the color
selection test; their number is given in parentheses.~~
(e)~Statistical excess of objects in the M31 halo field CMD in~(a)
over that in~(b).~~ (f)~Same as (e) corrected for excess foreground
contamination in the M31 halo field [(c) minus (d)].  \label{fig8}}
\end{figure}

We use the color- and morphology-selected, area-matched comparison field data
set [Fig.~\ref{fig8}(b)] to assess the amount of contamination
in the corresponding M31 halo field sample.  Each object in the comparison
field CMD is matched to the object ``closest'' to it in color-magnitude
location in the scaled M31 halo CMD, provided it is located within a local
elliptical region.  This elliptical region is arbitrarily defined by semi
axes lengths of 0.3~mag in $B-I$ color and 0.5~mag in apparent $I$ magnitude.
The matched pairs are removed from both CMDs.  Only 81~objects remain
unmatched out of the 757 in the $UBRI$- and FWHM-selected comparison field
sample ($\sim10$\%), while 1402 out of 2078 (67\%) remain unmatched in the
scaled M31 halo field.  This preliminary sample of ``excess'' objects in the
M31 halo field shows a strong concentration around the expected CMD location
of RGB stars at the distance of M31 [Fig.~\ref{fig8}(e)]: $\rm
CMD_{\ref{fig8}e} = CMD_{\ref{fig8}a} - CMD_{\ref{fig8}b}$.

The M31 halo field is closer to the Galactic plane ($\vert{b}\vert=22^\circ$) 
than the comparison field ($\vert{b}\vert=30^\circ$), and is expected to
contain 1.7~times more foreground Galactic dwarf stars than the area matched
comparison field sample on the basis of the IASG model [compare 
Figs.~\ref{fig8}(c) and (d)].  We carry out another statistical
subtraction between the two IASG model CMDs to estimate the number and
distribution in color-magnitude space of the {\it extra\/} foreground
Galactic stars in the scaled M31 halo field.  We also simulate the fall off
in $UBRI$ selection efficiency for low luminosity, red main sequence stars by
excluding an increasing fraction of stars with $B-I>3.5$; the excluded stars
are indicated by crosses in Figs.~\ref{fig8}(c) and (d).  This
sample of extra IASG model stars is then used to (statistically) remove {\it
residual\/} foreground contaminants from the preliminary
statistically-subtracted M31 halo field sample to obtain the final sample of
M31 halo RGB candidates [Fig.~\ref{fig8}(f)]: $\rm
CMD_{\ref{fig8}f} = CMD_{\ref{fig8}e} -
(CMD_{\ref{fig8}c} - CMD_{\ref{fig8}d})$.

\subsection{Star Selection Efficiency}
\begin{table}
\tablenum{1}
\begin{tabular}{lrrrrrrr}
\multicolumn{8}{c}{\smcap Table~1.} \\
\multicolumn{8}{c}{Fraction of Color and Morphology-selected Objects} \\
\hline
\hline
 
\multicolumn{1}{c} {$I$} &
\multicolumn{1}{c} {Full} &
\multicolumn{1}{c} {Color} &
\multicolumn{1}{c} {FWHM} &
\multicolumn{1}{c} {Combined} &
\multicolumn{1}{c} {Color/} &
\multicolumn{1}{c} {Combined/} &
\multicolumn{1}{c} {Combined/} \\

\multicolumn{1}{c} {} &
\multicolumn{1}{c} {Sample} &
\multicolumn{1}{c} {Selected} &
\multicolumn{1}{c} {Selected} &
\multicolumn{1}{c} {Selection} &
\multicolumn{1}{c} {Full} &
\multicolumn{1}{c} {Color} &
\multicolumn{1}{c} {Full} \\

\hline
\multicolumn{8}{c} {M31 Halo Field} \\
\hline
\hfill 17--18 &  144 &  102 &   75 &  55 & 0.71 & 0.54 & 0.38 \\
\hfill 18--19 &  192 &  136 &   89 &  64 & 0.71 & 0.47 & 0.33 \\
\hfill 19--20 &  237 &  130 &  104 &  66 & 0.55 & 0.51 & 0.28 \\
\hfill 20--21 &  430 &  223 &  176 & 101 & 0.52 & 0.45 & 0.23 \\
\hfill 21--22 &  805 &  379 &  314 & 184 & 0.47 & 0.49 & 0.23 \\
\hfill 22--23 & 1384 &  629 &  592 & 333 & 0.45 & 0.53 & 0.24 \\
\hfill 23--24 & 2156 &  992 & 1069 & 543 & 0.46 & 0.55 & 0.25 \\
\hfill 24--25 & 1996 & 1052 & 1103 & 620 & 0.53 & 0.59 & 0.31 \\
\hline
\multicolumn{8}{c} {(Matched) Comparison Field} \\
\hline
\hfill 17--18 &   64 &  47 &  37 &  30 & 0.73 & 0.64 & 0.47 \\
\hfill 18--19 &  109 &  54 &  53 &  33 & 0.50 & 0.61 & 0.30 \\
\hfill 19--20 &  211 & 105 &  72 &  50 & 0.50 & 0.48 & 0.24 \\
\hfill 20--21 &  298 & 127 & 103 &  65 & 0.43 & 0.51 & 0.22 \\
\hfill 21--22 &  488 & 189 & 149 &  75 & 0.39 & 0.40 & 0.15 \\
\hfill 22--23 &  837 & 275 & 326 & 116 & 0.33 & 0.42 & 0.14 \\
\hfill 23--24 & 1262 & 373 & 594 & 186 & 0.30 & 0.50 & 0.15 \\
\hfill 24--25 &  880 & 287 & 442 & 135 & 0.33 & 0.47 & 0.15 \\

\hline

\end{tabular}
\end{table}

The efficiency of the $UBRI$- and FWHM-based star selection procedure may be
estimated by studying the fraction of objects selected at every step of the
procedure, coupled with {\it a~priori\/} knowledge of the ratio of stars to
galaxies, as a function of apparent brightness.  Table~1 shows the selected
numbers and fractions of objects for both M31 halo and comparison fields in
1~mag bins in the $I$ band.  At bright magnitudes ($I<19$), the surface
density of field galaxies is only 0.43~arcmin$^2$ (Lilly et~al.\ 1995a,b) or
about 90~galaxies in the KPNO field, and foreground Galactic stars are
expected to outnumber field galaxies 7:2 in the low latitude M31 halo field.
Color selection appears to retain about two-thirds of all stars in the
brightest magnitude bins.  This is the expected fraction enclosed within the
$1.5\sigma$ ellipse (see $UBRI$ selection criterion in Sec.~3.2) for a two
dimensional Gaussian distribution.  Morphological selection picks out
$\ga50$\% of the remaining objects.  This fraction is significantly less than
unity (even at bright magnitudes where the effect of galaxy contamination is
negligible) because of measurement error in the angular size (due to photon
noise, crowding, etc.) coupled with the fact that the PSF quality is not
constant over the CCD image.  While increasing the value of $\theta_{\rm
crit}$ would include more stars, it would also correspondingly increase the
number of barely resolved galaxies that pass FWHM selection.  Overall the
combined selection efficiency for stars at bright magnitudes is $\sim30$\%.

At magnitudes fainter than about $I=20$, the $UBRI$ selection step picks
only $\sim50$\% of all objects.  It should be noted though that this is {\it
not\/} a direct measure of the star selection efficiency.  The measured
fraction of selected objects is decreased (relative to the `true' $UBRI$
selection efficiency for stars) by the presence of field galaxies in the
unselected sample which dominate the sample at the faintest magnitudes.  Some
of these galaxies happen to have star-like colors, however, and these
contaminants tend to increase the color-selected fraction.  The increased
importance of systematic, non-Gaussian errors at faint magnitudes (e.g.~due
to interference from close neighbors, sky background measurement error)
likely causes more than a third of the stars to escape color selection.  FWHM
selection at faint magnitudes retains $\sim50$\% of the color-selected
subsample, but this is not necessarily a good measure of the morphological
selection efficiency for stars for reasons explained above.  The true FWHM
selection efficiency at faint magnitudes is expected to be lower than 50\%
(the estimated efficiency at bright magnitudes) because of increased error in
the measurement of $\theta_{\rm FWHM}$.  Thus, the overall selection
efficiency for faint stars is likely in the range 10\%$\>$--$\>$20\%.

The comparison field data are also useful for investigating the selection
efficiency.  As noted earlier, this field is at a slightly higher Galactic
latitude than the M31 halo field and therefore contains a smaller number of
foreground Galactic dwarf stars.  Thus the star-to-galaxy ratio is lower in
the comparison field even for magnitudes brighter than $I=20$, a range in
which there is no contribution from M31 red giant stars in the M31 halo
field sample.  It is therefore no surprise that the fraction of color- and
morphology-selected objects is somewhat lower in the comparison field than in
the M31 halo field over the full range of apparent brightnesses (Table~1).  A
more detailed comparison between the selection efficiencies in the M31 halo
and comparison fields is precluded by subtle differences in their faint end
noise characteristics and in the seeing FWHM (which affect the $UBRI$- and
FWHM-selected fractions, respectively).

\section{Discussion}
\subsection{Estimating the Metallicity of M31's Outer Spheroid}

The metallicity of the population of M31 halo stars detected is estimated by
comparing the color-magnitude distribution of the $UBRI$- and FWHM-selected,
statistically-subtracted sample of objects to the RGB fiducials for the
Galactic globular clusters M5 and 47~Tuc in Fig.~\ref{fig8}(f).
Clearly the fiducial for M5 is not a good match as most of the M31 halo field
objects brighter than $I=23$ are redder, and hence more metal rich, than M5
($\rm [Fe/H]=-1.4$).  The clump of faint ($I>23$), blue ($B-I\sim1\>$--$\>$2)
objects in this plot is produced at least in part by residual contamination
by distant faint blue field galaxies [compare
Figs.~\ref{fig8}(c) and (f)].  The fiducial for 47~Tuc is a
better match to the data, and yet there is a substantial spread about this
fiducial.  A substantial fraction of the objects with $I<23$ lie to the right
of the 47~Tuc fiducial which indicates that the mean metallicity of M31's
outer halo is comparable to, and possibly a bit higher than, that of 47~Tuc
($\rm [Fe/H]=-0.70$), with a fairly large spread in color.  

The mean photometric errors estimated from Poisson noise are
$\sigma_{B-I}\simeq0.02$ for objects near the tip of the M31 RGB
($I\sim20.5$) and $\sigma_{B-I}\simeq0.18$ near the limit of the sample
($I\sim24$).  The observed spread in color is too large to be accounted for
simply in terms of photometric error.  Despite our efforts to remove all
possible contaminants, some of the observed spread could be due to residual
contamination from foreground Galactic stars and background faint compact
galaxies.  The IASG model has not been thoroughly tested at faint magnitudes;
moreover, Poisson fluctuations in the field galaxy counts cause differences
between the comparison and M31 halo fields.  These possible contaminants
cannot be entirely responsible for the observed color spread however as there
is no reason why the contaminants should contribute primarily near the
observed CMD location of the M31 RGB.  In fact, compact galaxies contribute
primarily at faint magnitudes ($I\ga22$) and blue colors ($B-I\leq2$) as can
be seen by comparing the $UBRI$- and FWHM-selected comparison field sample in
Fig.~\ref{fig8}(b) with the IASG model prediction for this
field in Fig.~\ref{fig8}(d).  The uncertainty of the IASG model
at faint magnitudes is harder to quantify; some Galactic stars may still be
contaminating even the foreground subtracted CMD in
Fig.~\ref{fig8}(f).  We conclude that much of the spread 
observed in the RGB concentration is due to a true spread in the metallicity
of M31 halo stars.  Future followup spectroscopy will test this conclusion
(Sec.~4.3).

\subsection{Comparison of this Study to Recent HST Studies}

In this section, we compare the final, statistically-subtracted sample of RGB
candidates in the M31 halo CMD to those found in the {\it HST\/} studies of
Holland et~al.\ (1996) and Rich et~al.\ (1996).  Fig.~\ref{fig1} shows the
locations of all the fields in relation to M31's disk and halo (schematic
representation).  We adopt the authors' plausible hypothesis that the vast
majority of objects seen in Rich et~al.'s $R=40$~kpc (G1)
CMD and in Holland et~al.'s $R=7.6$~kpc (G302) and 10.8~kpc (G312) CMDs are
M31 field halo stars (see details below).  We use the External Galaxy Model
(EGM) developed by Hodder (1995) to estimate the number of halo and disk
stars expected to be seen in each of these fields.  The EGM is an adaptation
of the Bahcall \& Soneira (1984) Galaxy model designed to give star counts
and color distributions for any field in an external spiral galaxy using a 
de~Vaucouleurs profile for the halo and an exponential disk. 
\begin{table}
\tablenum{2}
\begin{tabular}{lrrrrrrr}
\multicolumn{8}{c}{\smcap Table~2.} \\
\multicolumn{8}{c}{Observed Star Counts and EGM Predictions} \\
\hline
\hline
 
\multicolumn{1}{c} {Field} &
\multicolumn{1}{c} {Area} &
\multicolumn{1}{c} {$X^a$} &
\multicolumn{1}{c} {$Y^a$} &
\multicolumn{2}{c} {Observed} &
\multicolumn{2}{c} {EGM Predictions$^b$} \\

\multicolumn{1}{c} {} &
\multicolumn{1}{c} {[arcmin$^2$]} &
\multicolumn{1}{c} {[kpc]} &
\multicolumn{1}{c} {[kpc]} &
\multicolumn{1}{c} {$I$=20--22} &
\multicolumn{1}{c} {$I$=22--24} &
\multicolumn{1}{c} {$I$=20--22} &
\multicolumn{1}{c} {$I$=22--24} \\

\hline

This Study$\,\>$ \hfill & 210.0 & \hfill $-1.7$ & \hfill $-18.7$ & \hfill 122$^c
$ &
\hfill 543$^c$ & \hfill 421 (421) & \hfill 6141 (6140) \\
G302$^d$ \hfill & \hfill 3.3 & \hfill $-2.5$ & \hfill $-6.7$ & \hfill 339$\,\>$ 
&
\hfill 1756~  & \hfill 132$\,\>$(119) & \hfill 1955 (1748)\\
G312$^d$ \hfill & \hfill 3.3 & \hfill $+0.5$ & \hfill $-11.1$ & \hfill 82~$\,$ &
\hfill 436$\,\,$ & \hfill 33$\,\>$(32)$~\,$ & \hfill 486 (479)$~\,$ \\
G1$^e$ \hfill & \hfill 1.1 & \hfill $+33.6$  & \hfill $+4.6$ & \hfill 33~$\,$ &
\hfill 140$\,\,$ & \hfill 10$\,\,$(6)~~~ & \hfill 157$~$(91)~~~ \\

\hline

\end{tabular}
 
\begin{itemize}
\item[$^a$]{($X$, $Y$)~=~Projected distance of field center from M31's center
            along its major and minor axes, respectively}
\item[$^b$]{Numbers in parentheses indicate spheroid stars}
\item[$^c$]{The star selection efficiency for our $R=19$~kpc sample is
            estimated to be about 30\% for the brighter apparent magnitude
            range and probably as low as 10\%$\>$--$\>$20\% for fainter stars}
\item[$^d$]{From the study by Holland et~al.\ (1996)}
\item[$^e$]{From the study by Rich et~al.\ (1996)}

\end{itemize}
\end{table}

There is a large number of free parameters in the EGM (disk and halo scale
lengths, density ratio, halo axis ratio, and overall normalization), but a
plausible model based on 47~Tuc's red giant luminosity function and color
distribution, a disk exponential scale length of 5.5~kpc, a halo effective
radius of 2.6~kpc, and an axis ratio for the halo of~0.6 can be normalized to
match the star counts observed in the {\it HST\/} fields to within $\sim$3\%.
Furthermore, the fraction of M31 disk stars predicted by the EGM for the two
Holland et~al.\ fields agrees with the upper limits to the fractional disk
contamination estimated by the authors: 0.1 for G302 and 0.03 for G312.  The
halo effective radius of 2.6~kpc for the major axis corresponds to a minor
axis effective radius of 1.6~kpc for the flattened halo.  This is consistent
with the minor axis effective radius of 1.3~kpc found by Pritchet \&
van~den~Bergh (1994).  The EGM predicts about 6500~RGB stars with $20<I<24$
within the 210~arcmin$^2$ effective area of the KPNO image of the M31 halo.
Most of these are expected to be halo giants with negligible contamination
from M31's disk (0.01\%).   Table~2 presents the observed number of stars and
EGM predictions for this study and for the three {\it HST\/} fields.

The surface density of M31 stars predicted by the EGM in the $R=19$~kpc field
we have studied is lower than in any of the three {\it HST\/} fields, even
Rich et~al.'s G1 field at $R=40$~kpc.  A 5:3 flattened halo contributes about
equally to the G1 field and ours, but M31's disk contributes an additional
70\% (40\% of total) in the G1 field.  Contrary to the impression one may get
from Fig.~\ref{fig1}, the degree of disk contamination is much higher in
G1 than in the inner Holland et~al.\ (G302) field.  Even though the G302
field is expected to contain {\it more\/} disk stars than G1, this is more
than compensated by the much larger number of halo stars in G302, making the
fractional disk contamination in G302 quite small.  This is simply a
reflection of the fact that galaxy halos have steeper density profiles than
the typical disk exponential law.  The M31 disk RGB fraction in G1 could
be even higher if the halo is less flattened ($b/a>0.6$) and/or if a thick
disk is present.  Thus, the large spread in the M31 halo metallicity inferred
by Rich et~al.\ may in fact be due primarily to disk contamination (possibly
including stars with super-solar [Fe/H]). 

After FWHM and $UBRI$ selection and statistical subtraction, we detect 
665~objects down to a limit of $I<24$ in the M31 halo field [see
Fig.~\ref{fig8}(f)], whereas the EGM predicts about 6500.  This
would seem to indicate a $\sim10\%$ selection efficiency, but the efficiency
is higher than this over most of the magnitude range.  About $\sim30$\% of
the number predicted by the EGM is observed in the range $20<I<22$, which
is close to estimated efficiency of the star selection process at relatively
bright magnitudes (as discussed in Sec.~3.6).  At fainter magnitudes the
selection efficiency drops for a variety of reasons: non-gaussian photometric
errors (effect of crowding on isophotal definition, sky measurement errors,
etc.) and increased FWHM measurement error (causing stars to scatter beyond
$\theta_{\rm crit}$).  Overall, the number of objects seen in our field is
roughly consistent with the EGM model prediction given the selection
efficiency.

The estimated metallicity of M31's outer halo from our study (Sec.~4.1) is
consistent with that found in the recent {\it HST\/} studies.
Holland et~al.\ (1996) find a spread in metallicity of
$\rm-2\la[Fe/H]\la-0.2$ with the majority of stars having [Fe/H]$\simeq-0.6$.
The contamination due to M31 disk stars is estimated using the EGM to be
$\sim0.1$ for the G302 field and $\sim0.03$ for the G312 field.  This, along
with the estimated foreground Galactic contamination of only 15~disk stars,
cannot completely explain the population of very red (high metallicity)
objects observed in the G302 and G312 fields.  If the disk of M31 is warped,
however, the fractional disk contamination could be higher in these Holland
et~al.\ fields.  Rich et~al.\ (1996) find the field population around G1 to
be as metal rich as 47~Tuc, but warn that the apparent high metallicity could
also be due to the sample containing objects of intermediate age.  As
discussed above, contamination by M31's disk is expected to be significant in
the G1 field.

\subsection{Density and Size of the M31 Halo}

The observed RGB counts in the Holland et~al.\ (1996) {\it HST\/} minor axis
fields and in our $R=19$~kpc minor axis field can be used to constrain the
physical parameters of M31's halo.  The halo is assumed to be oblate and
viewed edge on---i.e.,~with its symmetry axis in the plane of the sky.  A
(deprojected) power law index of $\nu=-3.8$ fits the surface densities
observed in the $R=7.6$~kpc (G302) and $10.8$~kpc (G312) Holland et~al.\
fields derived from stars with $20<I<23.5$.  This slope is a bit shallower
than but consistent with the range of $\nu=-4$ to $-5$ found by Pritchet \&
van~den~Bergh (1994) over the (minor axis) radial range of
$R=3\>$--$\>$20~kpc based on stars within 2~mag of the tip of the RGB.
Pritchet \& van~den~Bergh find that M31's halo density profile steepens with
increasing radius (in a log-log plot), so it is not surprising that a single
power law fit to the inner halo ($R<10.8$~kpc) is shallower than one that
includes the outer halo.

For a power law density profile, the column density at a projected radial
distance $R$ along the minor axis is given by:  

\begin{equation}
N=\int_{-\infty}^{\infty}n_0\left[\frac{x^{2}+(R/c)^{2}}{R_0^{2}}\right]^{\nu/2}dx
\end{equation}
\begin{equation}
N=n_0\left(\frac{R}{c}\right)\left(\frac{R_0\,c}{R}\right)^{-\nu}G(\nu)
\end{equation}

\noindent
where $c=0.6$ is the halo axial ratio, $x$ is the distance along the line of
sight, and

\begin{equation}
G(\nu)\equiv\frac{\Gamma(-\nu/2\>-\>1/2)\,\Gamma(1/2)}{\Gamma(-\nu/2)} ~~~.
\end{equation}

\noindent
The volume density $n_0$, at a 
radial distance $R_0$ in the equatorial plane of the halo, may thus be
inferred from the observed surface density:

\begin{equation}
n_0={{\frac{N\,c}{R\,G(\nu)} \left(\frac{R}{R_0\,c}\right)^{-\nu}}}
\end{equation}

The halo volume density is calculated in the apparent magnitude range
$20<I\leq23.5$ to avoid red clump stars and to directly compare to Morrison's
(1993) measurement of the local Galactic halo stellar density.  A projected
density of $N=7870$~kpc$^{-2}$ is derived from counts of stars (1290) in this
$I$ magnitude range in the inner Holland et~al.\ field (G302).  The area of
this field (two~Wide Field Camera CCDs) is 3.3~arcmin$^2$ which corresponds
to 0.17~kpc$^2$ at the distance of M31.  The inner (7.6~kpc) Holland et~al.\
field, G302, is not exactly on the minor axis; the ellipse with $c=0.6$ that
intersects this field has a semi-minor axis length of $R=6.8$~kpc, and this
is the value used in the above equations.

As M31 is about $\Lambda\approx1.5$~times larger than the Galaxy, we choose
$R_0=12$~kpc in order to calculate the halo density at a position in M31
comparable to the solar location in the Galaxy.  The derived M31 halo density
at $R_0=12$~kpc is $n_0=341$~kpc$^{-3}$.  Thus, $n_0$ is much larger than the
local Galactic halo density of 36~kpc$^{-3}$ for stars with $M_V\leq0.5$
(Morrison 1993).  This indicates that M31's halo is much denser, and/or
larger than that of the Galaxy, $(\rho_{\rm M31}^{\rm RGB}/\rho_{\rm
MW}^{\rm RGB})(\Lambda/1.5)^{-\nu}\sim9.5\pm0.3$, where $\Lambda$ is the
ratio of the characteristic radial scale lengths of M31 and the Galaxy,
$\nu\sim-3.8$ is the power law density profile slope, and the (formal) error
bar is based on Poisson error in the star counts in the inner Holland
et~al.\ field, G302.  The uncertainty in the overdensity estimate is
substantially higher than the formal error bar quoted above when one takes
into account the possible range of profile slopes, $\nu$ (see below).

The actual density of M31's inner halo is likely to be even higher than this
estimate:  We have adopted a single power law profile in the above integral,
with a slope of $\nu=-3.8$ based on M31's inner halo, but the outer profile
is known to steepen both from the counts in our $R=19$~kpc field and from
Pritchet \& van~den~Bergh (1994); this leads to an overestimate of the column
density for a given spatial density $n_0$, or to an underestimate of $n_0$
since we normalize to the {\it observed\/} column density.  At any rate, it
is clear than the halo of M31 is significantly denser/larger than the halo of
the Galaxy.

The density profile of the outer halo of M31 (beyond a minor axis radius of
10~kpc) has a slope of $\nu\approx-5$, as indicated by the EGM de~Vaucouleurs
spheroid fit to the observed counts at $R=10.8$~kpc (G312, Holland et~al.\
1996) and $R=19$~kpc (this study), and by Pritchet \& van~den~Bergh's (1994)
study.  This is steeper than the profile of outer Galactic halo which may be
approximated by a power law index in the range $\nu_{\rm MW}=-3.0$ to $-3.5$
(Zinn 1985; Preston et~al.\ 1991; Gould et~al.\ 1998).  There are significant
uncertainties though in these measurements of the profile of the Galaxy's
outer stellar halo, and the halo of M31 is not well fit by a power law but
rather by a de~Vaucouleurs law (Pritchet \& van~den~Bergh 1994), so directly
comparing the power law indices may be misleading. 

\subsection{Future Work/Followup Spectroscopy}

Although it is becoming increasingly clear that the halo of M31 is fairly
metal rich, an accurate mean metallicity and spread remain to be determined.
Contamination from foreground Galactic stars and background field galaxies,
as well as from disk RGB stars in M31, is still non-negligible.  The combined
color- and morphological-selection technique described in this paper can be
used effectively to study M31's inner halo where the fractional contamination
is expected to be lower than in our $R=19$~kpc field.  High angular
resolution imaging (e.g.,~with {\it HST\/}) can help reduce background galaxy
contamination.

We are currently undertaking followup Keck/LRIS spectroscopy of the brightest
decade of M31 halo RGB candidates in the $UBRI$- and FWHM-selected sample
[Fig.~\ref{fig6}(d)] (Guhathakurta \& Reitzel 1998).  Candidates
for spectroscopic observation are selected without regard to color, so as not
to introduce any metallicity bias.  We do not select objects from the
statistically-subtracted version of the CMD as it is not complete.
Spectroscopy should allow kinematic confirmation of the identity of M31 RGB
stars: galaxies are eliminated altogether on the basis of their redshift and
M31 stars are quite well separated from foreground Galactic disk stars
($v^{\rm sys}_{\rm M31}\sim-300$~km~s$^{-1}$ vs $v_{\rm
MW~disk}\approx0$~km~s$^{-1}$).  A broad versus narrow distribution of radial
velocities will distinguish M31 halo stars from disk stars on M31's minor
axis, respectively.  In addition to removal of residual contaminants,
spectroscopy allows a direct measurement of the metallicity of each star
using the near infrared \Catwo\ line strengths, W$_{\rm Ca}$.  A well
developed empirical calibration method exists to determine [Fe/H] from
W$_{\rm Ca}$ and the brightness of the star $\delta{I}$ relative to the
horizontal branch level (Olszewski et~al.\ 1991; Armandroff \& Da Costa 1991;
Suntzeff \& Kraft 1996).

\section{Summary}

We analyze deep $UBRI$ images obtained with the KPNO 4-meter telescope of a
field in the outer halo of M31 ($R=19$~kpc in projection on the minor axis of
M31), and of a comparison field located at similar Galactic coordinates.
Keck/LRIS $I$ band images of the M31 halo field are used for the purpose of
making angular size measurements and morphological star-galaxy separation.
We use a $UBRI$ color selection technique in conjunction with morphological
selection to isolate a sample of M31 RGB stars from the numerous background
population of faint field galaxies.  A detailed comparison shows that the M31
halo field contains an excess of objects over the populations present in the
comparison field.  An empirical Galactic (IASG) model is used to estimate the
foreground contamination from Galactic stars in both fields; the predicted
number of bright stars agrees well with the observed distributions in the M31
halo field and the comparison field.  We use the comparison field data and
the IASG model to statistically subtract contaminants from the M31 halo
field, revealing a distribution of objects at the expected location of the
RGB at the distance of M31.  These stars have a mean metallicity of
$\rm[Fe/H]\geq-0.7$ with a significant spread about this value, ranging from
M5 ($\rm[Fe/H]=-1.4$) to more metal rich than 47~Tuc ($\rm[Fe/H]=-0.70$).
This result is in good agreement with recent {\it HST\/} studies of the halo
of M31 by Holland et~al.\ (1996) and Rich et~al.\ (1996).  The spatial
density of M31's halo red giants is an order of magnitude greater than the
red giant density at a comparable radius in the Milky Way halo.
Alternatively, the halo of M31 has a characteristic size scale almost 3~times
larger than that of the Galactic halo.  The outer stellar halo of M31 has a
steeper slope ($\nu\sim-5$) than that of the Galaxy ($\nu=-3$ to $-3.5$).

\bigskip\bigskip
The early phases of this research were supported by a CalSpace grant and a
Divisional Seed grant at UC Santa Cruz.
A.\ G.\ was supported in part by NSF grant AST~9420746 and in part by NASA
grant NAG~5-3111.  We would like to thank the anonymous referee for helpful
suggestions that have led to an improvement of the paper.  Some of the data
presented herein were obtained at Kitt Peak National Observatory, which is
operated by Association of Universities for Research in Astronomy, Inc.,
while some of the data were obtained at the W.~M.~Keck Observatory, which is
operated as a scientific partnership among the California Institute of
Technology, the University of California, and the National Aeronautics and
Space Administration.  The W.~M.~Keck Observatory was made possible by the
generous financial support of the W.~M.~Keck Foundation.

\vfill\eject


\bigskip
\bigskip

\begin{references}

\reference Armandroff, T.~E. \& Da~Costa, G.~S.\ 1991, \aj, 101, 1329

\reference Bahcall, J.~N., \& Soneira, R.~M.\ 1984, ApJS, 55, 67

\reference Bertelli, G., Bressan, A., Chiosi, C., Fagotto, F., \& Nasi, E.\
1994, A\&AS, 106, 275

\reference Burstein, D., \& Heiles, C.\ 1982, ApJS, 54, 33

\reference Cardelli, J.~A., Clayton, G.~C., \& Mathis, J.~S.\ 1989, \apj,
345, 245

\reference Carney, B.~W., Aguilar, L., Latham D.~W., \& Laird J.~B.\ 1990,
\aj, 99, 201

\reference Christian, C.~A., \& Heasley, J.~N.\ 1991, \aj, 101, 848

\reference Couture, J., Racine, R., Harris, W.~E., \& Holland, S.\ 1995, \aj,
109, 2050

\reference Crotts, A.~P.~S.\ 1986, \aj, 92, 292

\reference Davidge, T.~J.\ 1993, \apj, 409, 190

\reference de Vaucouleurs, G.\ 1958, \apj, 128, 465

\reference Durrell, P.~R., Harris, W.~E., \& Pritchet, C.~J.\ 1994, \aj, 108,
2114

\reference Eggen, O.~J., Lynden-Bell, D., \& Sandage, A.~R.\ 1962,
\apj, 136, 748

\reference Freedman, W.~F., \& Madore, B.~F.\ 1990, \apj, 365, 186 

\reference Gould, A., Bahcall, J.~N., \& Flynn, C.\ 1996, \apj, 465, 759

\reference Gould, A., Bahcall, J.~N., \& Flynn, C.\ 1997, \apj, 482, 913

\reference Gould, A., Flynn, C., \& Bahcall, J.~N.\ 1998, \apj, in press
(astro-ph~9711263)

\reference Gould, A., Guhathakurta, P., Richstone, D., \& Flynn, C.\ 1992,
\apj, 388, 345 (GGRF)

\reference Green, E.~M., Demarque, P., \& King, C.~R.\ 1987, The Revised Yale
Isochrones and Luminosity Functions (New Haven: Yale University Observatory)

\reference Guhathakurta, P., \& Reitzel, D.~B.\ 1998, ApJL, in preparation

\reference Hodder, P.~J.~C. 1995, Ph.D. thesis, University of British
Columbia 

\reference Holland, S., Fahlman, G.~G., \& Richer, H.~B.\ 1996, \aj, 112, 1035

\reference Holtzman, J.~A., Burrows, C.~J., Casertano, S., Hester, J.~J.,
Trauger, J.~T., Watson, A.~M., \& Worthey, G.\ 1995, \pasp, 107, 1065

\reference Huchra, J.~P., Brodie, J.~P., \& Kent, S.~M.\ 1991, \apj, 370, 495

\reference Jarvis, J.~F., \& Tyson, J.~A.\ 1981, \aj, 86, 476

\reference Landolt, A.~U.\ 1983, \aj, 88, 439  

\reference Larson, R.~B.\ 1974, \mnras, 166, 585  

\reference Lilly, S.~J., Hammer, F., Le~F\'evre, O., \& Crampton, D.\ 1995a,
\apj, 455, 75

\reference Lilly, S.~J., Le~F\'evre, O., Crampton, D., Hammer, F., \& Tresse,
L.\ 1995b, \apj, 455, 50

\reference Morrison, H.~L.\ 1993, \aj, 106, 578

\reference Mould, J.\ 1986, in Stellar Populations, edited by C.~Norman,
A.~Renzini, and M.~Tosi (Cambridge: Cambridge University Press), p.~9

\reference Mould, J., \& Kristian, J.\ 1986, \apj, 305, 59

\reference Oke, J.~B., Cohen, J.~G., Carr, M., Cromer, J., Dingizian, A.,
Harris, F.~H., Labrecque, S., Lucinio, R., Schall, W., Epps H., \& Miller,
J.\ 1995, \pasp, 107, 375

\reference Olszewski, E.~W., Schommer, R.~A., Suntzeff, N.~B., \& Harris,
H.~C.\ 1991, \aj, 101, 515

\reference Preston, G.~W., Shectman, S.~A., \& Beers, T.~C.\ 1991,
\apj, 375, 121

\reference Pritchet, C.~J., \& van den Bergh, S.\ 1988, \apj, 331, 135

\reference Pritchet, C.~J., \& van den Bergh, S.\ 1994, \aj, 107, 1730

\reference Ratnatunga, K.~U., \& Bahcall, J.~N.\ 1985, ApJS, 59, 63

\reference Reitzel, D.~B., Guhathakurta, P., \& Gould, A.\ 1996a, in
Formation of the Galactic Halo....Inside and Out, edited by H.~Morrison and
A.~Sarajedini (ASP Conf.\ Series, No.~92), p.~540

\reference Reitzel, D., Guhathakurta, P., \& Gould, A.\ 1996b, in New Light
on Galaxy Evolution, edited by R.~Bender and R.~L.~Davies (Kluwer,
Dordrecht), p.~437


\reference Rich, R.~M., Mighell, K.~J., Freedman, W.~L., \& Neill, J.~D.\
1996, \aj, 111, 768

\reference Searle, L., \& Zinn, R.\ 1978, \apj, 225, 357  

\reference Stetson, P.~B.\ 1987, \pasp, 99, 191

\reference Stetson, P.~B.\ 1992, in Astronomical Data Analysis Software,
edited by D.~M.~Worrall, C.~Biemesderfer, and J.~Barnes (ASP Conf.\ Series,
No.~25), p.~297

\reference Suntzeff, N.~B., \& Kraft, R.~P.\ 1996, \aj, 111, 1913 


\reference Valdes, F.\ 1982, SPIE, 331, 465


\reference Williams, R.~E., Blacker, B., Dickinson, M., Dixon, W.~V.,
Ferguson, H.~C., Fruchter, A.~S., Giavalisco, M., Gilliland, R.~L., Heyer,
I., Katsanis, R., Levay, Z., Lucas, R.~A., McElroy, D.~B., Petro, L., \&
Postman, M.\ 1996, \aj, 112, 1335

\reference Zinn, R.\ 1985, \apj, 293, 424

\reference Zinn, R.\ 1993, in The Globular Cluster-Galaxy Connection, edited
by G.~H.~Smith and J.~P.~Brodie (ASP Conf.\ Series, No.~48), p.~38 

\end{references}
\end{document}